\documentclass[%
reprint,
superscriptaddress,
showpacs,
showkeys,
preprintnumbers,
nofootinbib,
nobibnotes,
amsmath,
prd,
]{revtex4-1}

\usepackage{graphicx}

\usepackage[
colorlinks,
citecolor=blue,
]{hyperref}

\usepackage{textcomp} 

\begin{document}

\preprint{CERN--EP--2017--137}

\title{Measurement of the $\pi K$ atom lifetime and the $\pi K$ scattering 
	length}

\author{B.~Adeva} 
\affiliation{Santiago de Compostela University}
\author{L.~Afanasyev} 
\affiliation{JINR, Dubna, Russia} 
\author{Y.~Allkofer} 
\affiliation{Zurich University, Switzerland} 
\author{C.~Amsler} 
\altaffiliation{Now at Stefan Meyer Institute for Subatomic Physics, Vienna,
	 Austria} 
\affiliation{Zurich University, Switzerland}
\author{A.~Anania} 
\affiliation{INFN, Sezione di Trieste and Messina University, Messina, Italy} 
\author{S.~Aogaki} 
\affiliation{IFIN-HH, National Institute for Physics and Nuclear Engineering,
	 Bucharest, Romania}
\author{A.~Benelli}
\affiliation{Czech Technical University in Prague, Prague, Czech Republic} 
\author{V.~Brekhovskikh} 
\affiliation{IHEP, Protvino, Russia} 
\author{T.~Cechak} 
\affiliation{Czech Technical University in Prague, Prague, Czech Republic} 
\author{M.~Chiba} 
\affiliation{Tokyo Metropolitan University, Japan} 
\author{P.~Chliapnikov} 
\affiliation{IHEP, Protvino, Russia} 
\author{D.~Drijard} 
\affiliation{CERN, Geneva, Switzerland}
\author{A.~Dudarev} 
\affiliation{JINR, Dubna, Russia} 
\author{D.~Dumitriu}
\affiliation{IFIN-HH, National Institute for Physics and Nuclear Engineering,
	 Bucharest, Romania} 
\author{P.~Federicova} 
\affiliation{Czech Technical University in Prague, Prague, Czech Republic}
\author{D.~Fluerasu} 
\affiliation{IFIN-HH, National Institute for Physics and Nuclear Engineering,
	 Bucharest, Romania} 
\author{A.~Gorin} 
\affiliation{IHEP, Protvino, Russia} 
\author{O.~Gorchakov} 
\affiliation{JINR, Dubna, Russia}
\author{K.~Gritsay} 
\affiliation{JINR, Dubna, Russia}
\author{C.~Guaraldo} 
\affiliation{INFN, Laboratori Nazionali di Frascati, Frascati, Italy} 
\author{M.~Gugiu} 
\affiliation{IFIN-HH, National Institute for Physics and Nuclear Engineering,
	 Bucharest, Romania} 
\author{M.~Hansroul} 
\affiliation{CERN, Geneva, Switzerland} 
\author{Z.~Hons} 
\affiliation{Nuclear Physics Institute ASCR, Rez, Czech Republic}
\author{S.~Horikawa} 
\affiliation{Zurich University, Switzerland}  
\author{Y.~Iwashita} 
\affiliation{Kyoto University, Kyoto, Japan} 
\author{V.~Karpukhin} 
\affiliation{JINR, Dubna, Russia} 
\author{J.~Kluson} 
\affiliation{Czech Technical University, Prague, Czech Republic} 
\author{M.~Kobayashi} 
\affiliation{KEK, Tsukuba, Japan} 
\author{V.~Kruglov} 
\affiliation{JINR, Dubna, Russia} 
\author{L.~Kruglova} 
\affiliation{JINR, Dubna, Russia} 
\author{A.~Kulikov} 
\affiliation{JINR, Dubna, Russia} 
\author{E.~Kulish} 
\affiliation{JINR, Dubna, Russia}  
\author{A.~Kuptsov} 
\affiliation{JINR, Dubna, Russia} 
\author{A.~Lamberto} 
\affiliation{INFN, Sezione di Trieste and Messina University, Messina, Italy} 
\author{A.~Lanaro} 
\affiliation{University of Wisconsin, Madison, USA} 
\author{R.~Lednicky} 
\affiliation{Institute of Physics ASCR, Prague, Czech Republic} 
\author{C.~Mari\~nas} 
\affiliation{Santiago de Compostela University}
\author{J.~Martincik} 
\affiliation{Czech Technical University, Prague, Czech Republic} 
\author{L.~Nemenov} 
\affiliation{JINR, Dubna, Russia} 
\affiliation{CERN, Geneva, Switzerland} 
\author{M.~Nikitin} 
\affiliation{JINR, Dubna, Russia} 
\author{K.~Okada} 
\affiliation{Kyoto Sangyo University, Japan} 
\author{V.~Olchevskii} 
\affiliation{JINR, Dubna, Russia} 
\author{M.~Pentia} 
\affiliation{IFIN-HH, National Institute for Physics and Nuclear Engineering,
	 Bucharest, Romania} 
\author{A.~Penzo} 
\affiliation{INFN, Sezione di Trieste, Trieste, Italy} 
\author{M.~Plo} 
\affiliation{Santiago de Compostela University}
\author{P.~Prusa} 
\affiliation{Czech Technical University, Prague, Czech Republic}   
\author{G.~Rappazzo} 
\affiliation{INFN, Sezione di Trieste and Messina University, Messina, Italy}  
\author{A.~Romero Vidal}
\affiliation{INFN, Laboratori Nazionali di Frascati, Frascati, Italy}   
\author{A.~Ryazantsev} 
\affiliation{IHEP, Protvino, Russia} 
\author{V.~Rykalin} 
\affiliation{IHEP, Protvino, Russia}  
\author{J.~Saborido} 
\affiliation{Santiago de Compostela University}
\author{J.~Schacher} 
\affiliation{Albert Einstein Center for Fundamental Physics, Laboratory of High
	 Energy Physics, Bern, Switzerland}  
\author{A.~Sidorov} 
\affiliation{IHEP, Protvino, Russia}  
\author{J.~Smolik} 
\affiliation{Czech Technical University in Prague, Prague, Czech Republic}  
\author{F.~Takeutchi} 
\affiliation{Kyoto Sangyo University, Japan}  
\author{L.~Tauscher}
\affiliation{Basel University, Switzerland} 
\author{T.~Trojek} 
\affiliation{Czech Technical University in Prague, Prague, Czech Republic} 
\author{S.~Trusov} 
\affiliation{Skobeltsyn Institute for Nuclear Physics of Moscow State 
	University, Russia} 
\author{T.~Urban}
\affiliation{Czech Technical University in Prague, Prague, Czech Republic} 
\author{T.~Vrba}
\affiliation{Czech Technical University in Prague, Prague, Czech Republic} 
\author{V.~Yazkov} 
\affiliation{Skobeltsyn Institute for Nuclear Physics of Moscow State 
	University, Russia} 
\author{Y.~Yoshimura} 
\affiliation{KEK, Tsukuba, Japan} 
\author{M.~Zhabitsky} 
\affiliation{JINR, Dubna, Russia} 
\author{P.~Zrelov} 
\affiliation{JINR, Dubna, Russia}

\collaboration{DIRAC Collaboration}
\noaffiliation

\date{\today}

\begin{abstract}
After having announced the statistically significant observation (5.6~$\sigma$)
of the new exotic $\pi K$ atom, the DIRAC experiment at the CERN proton 
synchrotron presents the measurement of the corresponding atom lifetime, based on 
the full $\pi K$ data sample: $\tau = (5.5^{+5.0}_{-2.8}) \cdot 10^{-15}s$. 
By means of a precise relation ($<1\%$) between atom lifetime and 
scattering length, the following value for 
the S-wave isospin-odd $\pi K$ scattering length  
$a_0^{-}~=~\frac{1}{3}(a_{1/2}-a_{3/2})$ has been derived:
$\left|a_0^-\right| = (0.072^{+0.031}_{-0.020}) M_{\pi}^{-1}$. 
\end{abstract}



\maketitle

\section{Introduction}

In 2007, the DIRAC collaboration enlarged the scope of 
the dimesonic atom investigation by starting to search for 
the strange pion-kaon ($\pi K$) atom. In addition to the ongoing study of 
$\pi\pi$ atoms, the DIRAC experiment at the CERN proton synchrotron (CERN PS) 
also collected data containing a kaon beside a pion in the final state. 
Using all the data since 2007 and optimizing data handling and analysis, 
the observation of the $\pi K$ atom could be achieved for the 
first time with a significance of more than 5 standard deviations \cite{ADEV16}. 
On the basis of the same data sample, this paper presents 
the resulting $\pi K$ atom lifetime and 
the corresponding $\pi K$ scattering length.

Using non-perturbative lattice QCD (LQCD), chiral perturbation theory (ChPT) 
and dispersive analysis, the S-wave $\pi\pi$ and $\pi K$ scattering lengths 
were calculated. S-wave $\pi\pi$ scattering lengths as described in QCD  
exploiting chiral $SU(2)_L\times SU(2)_R$ symmetry breaking were confirmed 
experimentally at a level of about 4\% \cite{BATE09,BATE10,ADEV11}. 
These measurements - independently of their accuracy - cannot test QCD 
predictions in the strange sector based on chiral $SU(3)_L\times SU(3)_R$ 
symmetry breaking. However, this check can be done by investigating 
$\pi K$ scattering lengths, where the s quark is involved. 

\begin{figure}[ht]
\begin{center}
\includegraphics[width=0.5\linewidth]{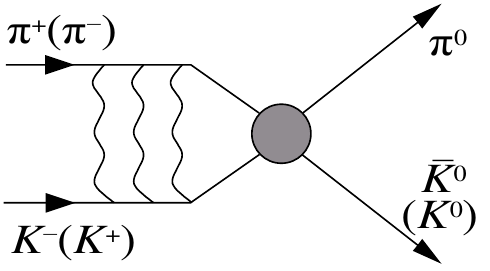}
\caption{The dominant decay channel of the $\pi K$ atom. 
The wavy lines indicate Coulomb photons. }
\label{fig:piK}
\end{center}
\end{figure}

The lifetime of the hydrogen-like $\pi K$ atom $A_{K \pi}$ or $A_{\pi K}$, 
consisting of $\pi^- K^+$ or  $\pi^+ K^-$ mesons, is given by 
the S-wave $\pi K$ scattering length difference 
$|a_{1/ 2}-a_{3/2}|$, where $a_I$ is the scattering length for 
isospin $I$ \cite{BILE69}. This atom is an electromagnetically bound state of 
$\pi^\mp$ and $K^\pm$ mesons with a Bohr radius of $a_{B}=249$~fm and 
a ground state Coulomb binding energy of $E_{B}=2.9$~keV. It decays 
predominantly\footnote{
Further decay channels with photons and $e^+ e^-$ pairs are suppressed at 
$\mathcal{O}(10^{-3})$.} 
by strong interaction into the neutral meson pair $\pi^0 K^0$ or 
$\pi^0 \overline{K}^0$ (Fig.~\ref{fig:piK}). 

The atom decay width $\Gamma_{1S}$ in the ground state (1S) is determined by 
the relation \cite{BILE69, SCHW04}:
\begin{align}
\label{eq:gamma}
\Gamma_{1S} & = \frac{1}{\tau_{1S}} \simeq 
\Gamma(A_{K \pi} \to \pi^0 K^0 \;\; \text{or} \;\; A_{\pi K} \to \pi^0
\overline{K}^0) \nonumber \\
&= 8 \; \alpha^3 \; \mu^2 \; p^* \; (a_{0}^-)^2 \; (1+\delta_K), 
\end{align}
where the S-wave isospin-odd $\pi K$ scattering length  
$a_0^-=\frac{1}{3}(a_{1/2}-a_{3/2})$ is defined in pure QCD for the quark masses 
$m_u=m_d$. Further, $\alpha$ is the fine structure constant, $\mu=109$~MeV/$c$ 
the reduced mass of the $\pi^{\mp} K^{\pm}$ system, 
$p^*=11.8$~MeV/$c$ the outgoing 3-momentum of $\pi^0$ or $K^0$
($\overline{K}^0$)
in the $\pi K$ atom system, and $\delta_K$ accounts 
for corrections, due to isospin breaking, at order $\alpha$ and 
quark mass difference $m_u - m_d$ \cite{SCHW04}.

A dispersion analysis of $\pi K$ scattering, using Roy-Steiner 
equations and 
experimental data in the GeV~range, yields 
$M_\pi (a_{1/2}-a_{3/2})=0.269\pm0.015$ \cite{BUET04}, 
with $M_\pi$ as charged pion mass. 
Inserting $a^-_0 = (0.090 \pm 0.005)~M_{\pi}^{-1}$ 
and $\delta_K = 0.040 \pm 0.022$ \cite{SCHW04} in (\ref{eq:gamma}), 
one predicts for the $\pi K$ atom lifetime in the ground state 
\begin{equation}
  \label{eq:tau35}
   \tau =(3.5 \pm 0.4)\cdot10^{-15}~\text{s}.
\end{equation}
In the framework of $SU(3)$ ChPT \cite{WEIN66,Gasser85}, $a_{1/2}$ and 
$a_{3/2}$ were calculated in leading order ($LO$) \cite{WEIN66}, 
1-loop ($1l$) \cite{BERN91} (see also \cite{KUBI02}) and 
2-loop order ($2l$) \cite{BIJN04}. This chiral expansion can 
be summarized as follows:
\begin{align}
  \label{eq:chiral}
   M_\pi a^-_0 & = M_\pi a^-_0(LO) (1+\delta_{1l}+\delta_{2l}) \nonumber \\
   & = M_\pi \frac{\mu}{8\pi F_\pi^2} (1+0.11+0.14) = 0.089
\end{align}
with the physical pion decay constant $F_\pi$, the 1-loop $\delta_{1l}$ 
and the 2-loop contribution $\delta_{2l}$. 
Because of the relatively large s quark mass, compared to 
u and d quark, chiral symmetry is much more broken, and ChPT 
is not very reliable at the $\pi K$ threshold. The hope 
is to get new insights by LQCD. Previously, $\pi K$~scattering lengths 
were investigated on the lattice with unphysical meson masses 
and then chirally extrapolated to the physical point. Nowadays, 
scattering lengths can be calculated directly at the physical point 
as presented in \cite{JANO14}: $M_\pi a^-_0 = 0.0745 \pm 0.0020$. 
Taking into account statistical and systematic errors, 
the different lattice calculations \cite{JANO14,BEAN06,Fu12,Sasaki14} 
provide consistent results for $a^-_0$. Hence, a scattering length 
measurement could sensitively check QCD (LQCD) predictions. 

\begin{figure}[ht]
\begin{center}
\includegraphics[width=\linewidth]{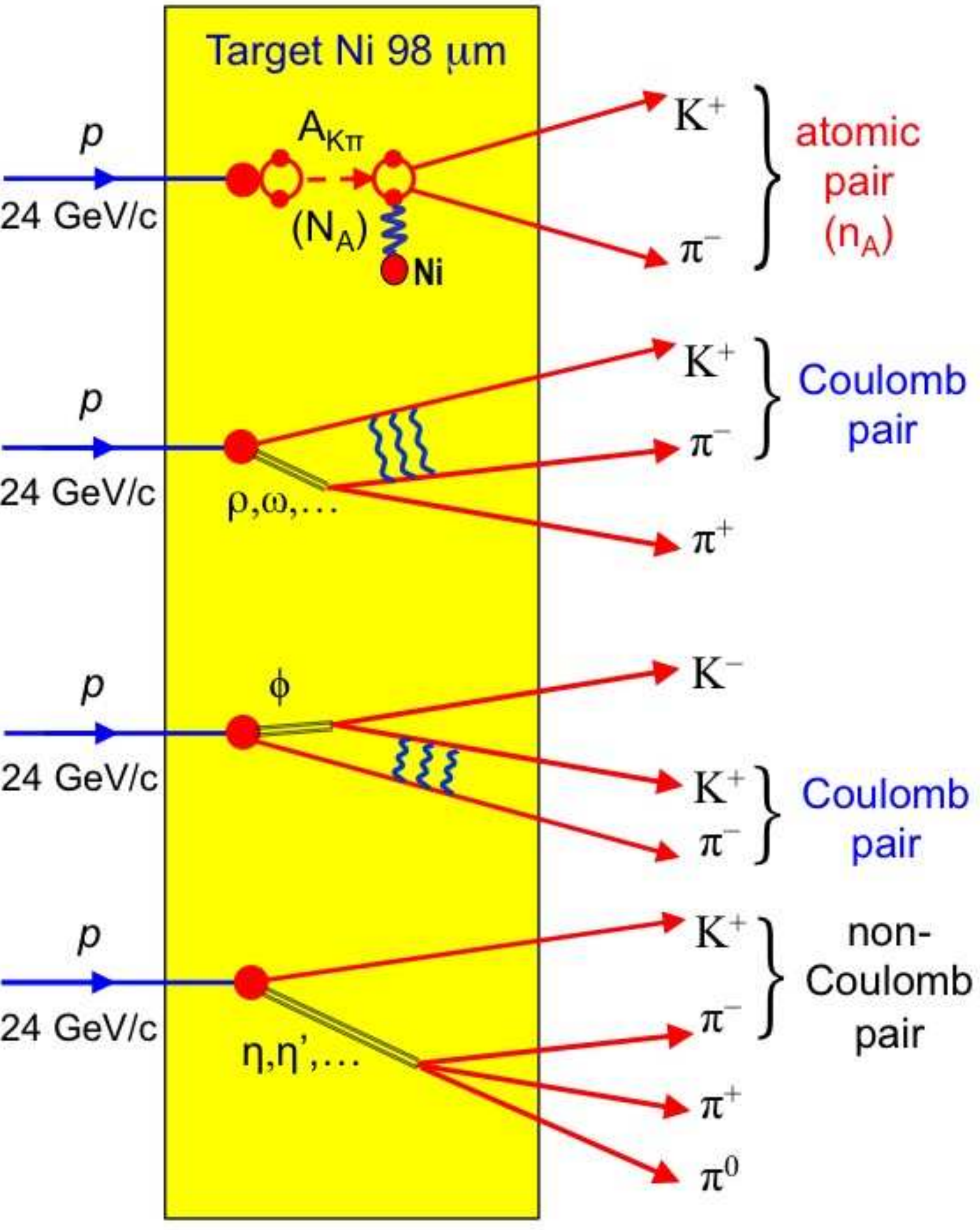}
\caption{Inclusive $\pi K$ production in the 
24~GeV$/c$ p-Ni interaction: 
p + Ni $\to$ $\pi^{\mp} K^{\pm}$ + X;~$A_{K \pi}$ stands for $K^+\pi^-$~atom. }
\label{fig:piK-source}
\end{center}
\end{figure}

The production of dimesonic atoms (mesonium) in 
inclusive high-energy interactions was described in 1985 \cite{NEME85}. 
To observe and study such atoms, the following sequence of physical steps 
was considered: 
production rate of atoms and their quantum numbers,
atom breakup by interacting electromagnetically with 
target atoms, lifetime measurement and background estimation. 
An approach to measure the lifetime, describing the atom as 
a multilevel system propagating and interacting in the target, 
was derived in \cite{afan96}. It provides a one-to-one relation 
between the atom lifetime and its breakup probability in the target. 
By this means, $\pi^+\pi^-$ \cite{AFAN93,AFAN94,ADEV04,ADEV05,ADEV11,ADEV15} 
and $\pi K$ atoms \cite{ADEV09,ADEV14,ADEV16} were detected and 
studied in detail by the DIRAC experiment. The $\pi K$ atom production in 
proton-nucleus collisions was calculated for different proton energies 
and atom emission angles \cite{GORC00,GORC16}. The relativistic 
$\pi K$ atoms, formed by Coulomb final state interaction (FSI), propagate 
inside a target and part of them break up (Fig.~\ref{fig:piK-source}). 
Particle pairs from breakup, called ``atomic pairs'' (atomic pair in
Fig.~\ref{fig:piK-source}), are characterised by small relative
momenta, $Q < 3$~MeV/$c$, in the centre-of-mass (c.m.) system of the
pair. Here, $Q$ stands for the experimental c.m. relative momentum, 
smeared by multiple scattering in the target and other materials and
by reconstruction uncertainties. Later, the original c.m. relative
momentum $q$ will also be used in the context of particle pair
production. In the small $Q$ region, the number of atomic pairs above 
a substantial background of free $\pi K$ pairs can be extracted.  

In the first $\pi K$ atom investigation with a platinum (Pt) target
\cite{ADEV09},
$173\pm54$ (3.2~$\sigma$) $\pi K$ atomic pairs were identified. 
This sample allowed to derive a lower limit on the $\pi K$ atom lifetime of 
$\tau > 0.8\cdot10^{-15}$~s (90\%~CL). For measuring the lifetime, 
a nickel (Ni) target was used because of its breakup probability 
rapidly rising with lifetime around $3.5\cdot10^{-15}$~s.  
This experiment yielded $178\pm49$ (3.6~$\sigma$) $\pi K$
atomic pairs, resulting in a first atom lifetime and a scattering length 
measurement \cite{ADEV14}: 
$\tau=(2.5^{+3.0}_{-1.8})\cdot10^{-15}$~s and $M_\pi a^-_0
=0.11^{+0.09}_{-0.04}$.
Next, the Pt and Ni data were reprocessed \cite{ADEV16} 
with more precise setup geometry, improved detector response 
description for the simulation and optimized criteria for 
the $\pi K$ atomic pair identification. The components of $Q_T$, 
the transverse component of $\vec{Q}$, are labelled $Q_X$ and 
$Q_Y$ (horizontal and vertical), and $Q_L$ is the longitudinal component. 
Concerning Pt data, informations from detectors upstream of 
the spectrometer magnet were included, improving significantly 
the resolution in $Q_T$ compared to the previous analyzis \cite{ADEV09}. 
By analyzing the reprocessed Pt and Ni data, 
$349\pm62$ (5.6~$\sigma$) $\pi^-K^+$ and $\pi^+K^-$ atomic pairs \cite{ADEV16} 
were observed with reliable statistics and the atom lifetime and 
scattering length measurement could be improved as presented here.

\section{Setup and conditions}
\label{sec:setup}

\begin{figure*}[ht]
\begin{center}
\includegraphics[width=\linewidth]{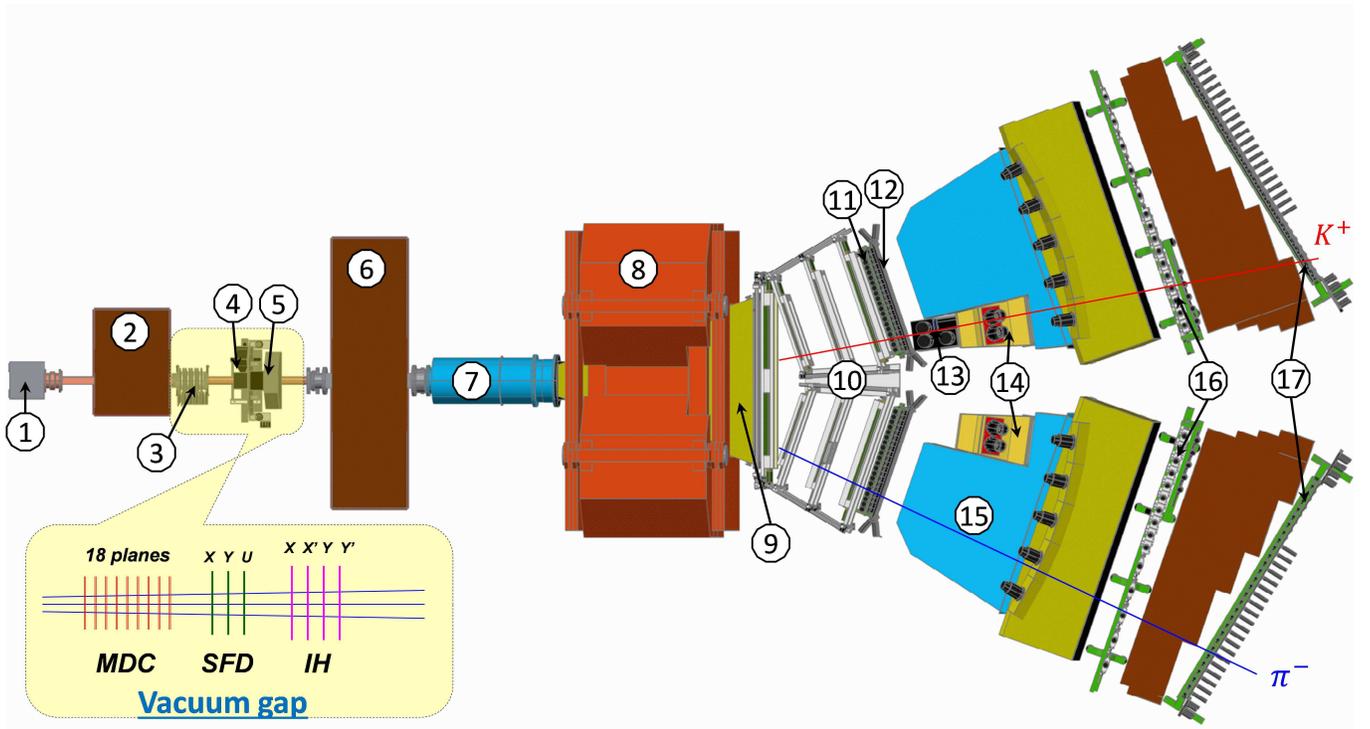}
\caption{ General view of the DIRAC setup 
(1 -- target station;
2 -- first shielding;
3 -- micro drift chambers (MDC);
4 -- scintillating fiber detector (SFD); 
5 -- ionization hodoscope (IH); 
6 -- second shielding; 
7 -- vacuum tube; 
8 -- spectrometer magnet; 
9 -- vacuum chamber; 
10 -- drift chambers (DC); 
11 -- vertical hodoscope (VH); 
12 -- horizontal hodoscope (HH); 
13 -- aerogel Cherenkov (ChA); 
14 -- heavy gas Cherenkov (ChF); 
15 -- nitrogen Cherenkov (ChN); 
16 -- preshower (PSh); 
17 -- muon detector (Mu).}
\label{fig:det}
\end{center}
\end{figure*}

The aim of the setup is to detect and identify simultaneously 
$\pi^- K^+$, $\pi^+ K^-$ and $\pi^+ \pi^-$ pairs with small $Q$. 
The magnetic 2-arm vacuum spectrometer \cite{DIRAC2} (Fig.~\ref{fig:det}) 
was optimized for simultaneous detection of 
these pairs \cite{GORC05a, GORC05b,PENT15}.
The structure of these pairs after the magnet is 
approximately symmetric for $\pi^+ \pi^-$ and asymmetric for $\pi K$ 
as sketched in Fig.~\ref{fig:det}. Originating from a bound system,  
these pair particles travel with similar velocities, and hence for 
$\pi K$ the K momentum is by the factor $\frac{M_{K}}{M_{\pi}}=3.5$ 
larger than the $\pi$ momentum, where $M_{K}$ is the charged kaon mass. 

The 24~GeV/$c$ primary proton beam, extracted from the CERN PS, 
hit in RUN1 a Pt target and in RUN2, RUN3 and RUN4 Ni~targets
(Table~\ref{tab:targ}).
The Ni targets are adapted for measuring the $\pi K$ atom lifetime, whereas 
the Pt target provides better conditions for the atom observation.
With a spill duration of 450~ms, the beam intensity was 
$(1.5 \div 2.1)\cdot10^{11}$ in 
RUN1 and $(1.05 \div 1.2)\cdot10^{11}$ protons/spill in RUN2 to RUN4, and the 
corresponding flux in the secondary channel $(5 \div 6) \cdot 10^6$
particles/spill.

\begin{table}[h]
\caption{Data and targets}
\label{tab:targ}
\begin{ruledtabular}
\begin{tabular}{p{0.258\columnwidth}cccc} 
Run Number & 1 & 2 & 3 & 4 \\ \hline 
Year & 2007 & 2008 & 2009 & 2010 \\ \hline
\raggedright Run duration & 3 months & 3 months & 5.3 months & 5.8 months\\
\hline
Target material & Pt & Ni & Ni & Ni \\ \hline
\raggedright Target purity (\%)& 99.95 & 99.98 & 99.98 & 99.98 \\ \hline
\raggedright Target thickness (\textmu{}m) & $25.7\pm1$ & $98\pm1$ & $108\pm1$ &
$108\pm1$ \\ \hline
Radiation thickness ($X_0$) & $8.4\cdot10^{-3}$ & $6.7\cdot10^{-3}$ &
$7.4\cdot10^{-3}$ & $7.4\cdot10^{-3}$ \\ \hline
\raggedright Nuclear efficiency & $2.8\cdot10^{-4}$ & $6.4\cdot10^{-4}$ &
$7.1\cdot10^{-4}$ & $7.1\cdot10^{-4}$ \\
\end{tabular}
\end{ruledtabular}
\end{table}

After the target station, primary protons pass under the setup 
to the beam dump, whereas secondary particles are confined by 
the rectangular beam collimator of the second steel shielding wall. 
The axis of the secondary channel is inclined relative to the 
proton beam by $5.7^\circ$ upward, and the angular divergence in the 
vertical and horizontal plane is $\pm 1^\circ$ (solid angle 
$\Omega = 1.2 \cdot 10^{-3}$ sr). Secondary particles propagate 
mainly in vacuum up to the Al foil $(7.6 \cdot 10^{-3} X_{0})$ 
at the exit of the vacuum chamber, which is installed between the  
poles of the dipole magnet ($B_{max}$ = 1.65~T and $BL$ = 2.2~Tm).

In the vacuum channel gap, 18 planes of the Micro Drift Chambers (MDC) 
and ($X$, $Y$, $U$) planes of the Scintillation Fiber Detector (SFD) 
were installed in order to measure both 
the particle coordinates  
($\sigma_{SFDx} = \sigma_{SFDy} = 60$~\textmu{}m, 
$\sigma_{SFDu} = 120$~\textmu{}m)
and the particle time 
($\sigma_{tSFDx} = 380$~\textmu{}m, 
$\sigma_{tSFDy} = \sigma_{tSFDu} = 520~\text{ps}$).
In RUN1 only the $Y$ and $U$ SFD planes were used. 
Four planes of the scintillation ionization hodoscope (IH) serve 
to identify unresolved double tracks (signal only from one SFD column). 
In RUN1 IH was not in use.
The total matter radiation thickness between target and vacuum chamber 
amounts to $7.7 \cdot 10^{-2} X_{0}$.

Each spectrometer arm is equipped with the following subdetectors \cite{DIRAC2}: 
drift chambers (DC) to measure particle coordinates with 
$\approx$85~\textmu{}m precision; 
vertical hodoscope (VH) to measure particle times with 110~ps accuracy 
to identify particle types via time-of-flight (TOF) measurement; 
horizontal hodoscope (HH) to select particles with a vertical distance of 
less than 75~mm ($Q_{Y}$ less than 15~MeV/$c$) in the two arms; 
aerogel Cherenkov counter (ChA) to distinguish kaons from protons; 
heavy gas ($\text{C}_{4} \text{F}_{10}$) Cherenkov counter (ChF) to 
distinguish pions from kaons; nitrogen Cherenkov (ChN) and 
preshower (PSh) counter to identify $\text{e}^+\text{e}^-$ pairs; 
iron absorber; 
two-layer muon scintillating counter (Mu) to identify muons. 
In the ``negative'' arm, no aerogel counter was installed, 
because the number of antiprotons compared to $K^{-}$ is small.

Pairs of oppositely charged time-correlated particles (prompt pairs) 
and accidentals in the time interval $\pm 20~\text{ns}$ are selected by 
requiring a 2-arm coincidence (ChN in anticoincidence) with 
the coplanarity restriction (HH) in the first-level trigger. 
The second-level trigger selects events with at least one track 
in each arm by exploiting the DC-wire information (track finder). 
Using the track information, the online trigger selects $\pi \pi$ and 
$\pi K$ pairs with relative momenta $|Q_X| < 12~\text{MeV}/c$ and 
$|Q_L| < 30~\text{MeV}/c$. The trigger efficiency is $\approx$ 98\% for 
pairs with $|Q_X| < 6~\text{MeV}/c$, $|Q_Y| < 4~\text{MeV}/c$ and 
$|Q_L| < 28~\text{MeV}/c$.
Particle pairs $\pi^{-} p$ ($\pi^{+} \bar{p}$) from 
$\Lambda$ ($\bar{\Lambda}$) decay were used for spectrometer calibration 
and $\text{e}^+\text{e}^-$ pairs for general detector calibrations.

\section{Production of bound and free $\pi^- K^+$ and $\pi^+ K^-$ pairs}
\label{sec:freebound}
Prompt oppositely charged $\pi K$ pairs, emerging from proton-nucleus
collisions, are produced either directly or originate from short-lived 
(e.g. $\Delta$, $\rho$),
medium-lived (e.g. $\omega$, $\phi$) or long-lived sources 
(e.g. $\eta'$, $\eta$).
These pion-kaon pairs, except those from long-lived sources, 
undergo Coulomb FSI resulting in modified unbound states 
(Coulomb pair in Fig.~\ref{fig:piK-source}) or forming bound systems in 
$S$-states with a known distribution of the principal quantum number 
($A_{K \pi}$ in Fig.~\ref{fig:piK-source}) \cite{NEME85}. 
Pairs from long-lived sources are nearly unaffected by the Coulomb interaction 
(non-Coulomb pair in Fig.~\ref{fig:piK-source}). The accidental pairs 
arise from different proton-nucleus interactions. 

The cross section of $\pi K$ atom production 
is given in \cite{NEME85} by the expression: 
\begin{align}
\label{eq:prod}
\frac{d\sigma^{n}_A}{d\vec p_A} & =(2\pi)^3\frac{E_A}{M_A}
\left.\frac{d^2\sigma^0_s}{d\vec p_K d\vec p_\pi}\right |_{
\frac{\vec p_K}{M_{K}} \approx \frac{\vec p_\pi}{M_{\pi}}}
\hspace{-1mm} \cdot \left|\psi_{n}(0)\right|^2 \nonumber \\
& = (2\pi)^3\frac{E_A}{M_A}
\frac{1}{\pi a_B^3 n^3}
\left.\frac{d^2\sigma^0_s}{d\vec p_K d\vec p_\pi}\right |_{
\frac{\vec p_K}{M_{K}} \approx \frac{\vec p_\pi}{M_{\pi}}} \:,
\end{align}
where $\vec p_{A}$, $E_{A}$ and $M_{A}$ are the momentum, 
energy and rest mass of the $A_{K \pi}$ atom in the 
laboratory system, respectively, and $\vec p_K$ and 
$\vec p_\pi$ the momenta of the charged kaon and pion 
with equal velocities. Therefore, these momenta obey 
in good approximation the relations 
$\vec p_{K}=\frac{M_{K}}{M_{A}} \vec p_{A}$ and 
$\vec p_{\pi}=\frac{M_{\pi}}{M_{A}} \vec p_{A}$. 
The inclusive production cross section of $\pi K$ pairs 
from short-lived sources without FSI is denoted by $\sigma_s^0$, 
and $\psi_{n}(0)$ is the $S$-state atomic Coulomb wave function 
at the origin with the principal quantum number $n$.  
According to (\ref{eq:prod}), $\pi K$ atoms are only produced in 
$S$-states with probabilities $W_n~=~\frac{W_1}{n^3}$:      
$W_1~=~83.2\%$, $W_2~=~10.4\%$, $W_3~=~3.1\%$, $\dots$ , $W_{n>3}~=~3.3\%$. 
In complete analogy, the production of free oppositely 
charged $\pi K$ pairs from short- and medium-lived sources,
i.e. Coulomb pairs, is described in the pointlike 
production approximation by 
\begin{align}
\label{eq:cross_sect_C}
\frac{d^2\sigma_C}{d\vec p_K d\vec p_\pi} & =
\frac{d^2\sigma^0_s}{d\vec p_K d\vec p_\pi}
\hspace{-2mm} \cdot A_C(q) 
\quad \text{with} \nonumber \\
A_C(q) & = \frac{2\pi m_\pi \alpha/q}
{1-\exp\left( -2\pi m_\pi \alpha/q\right) } \;.
\end{align}
The Coulomb enhancement function $A_C(q)$ in dependence on 
the relative momentum $q$ (see above) is the well-known 
Gamov-Sommerfeld-Sakharov factor \cite{GAMO28, SOMM31, SAKH91}. 
The relative yield between atoms and Coulomb pairs \cite{AFAN99} is 
given by the ratio of equations (\ref{eq:prod}) and (\ref{eq:cross_sect_C}). 
The total number $N_A$ of produced $A_{\pi K}$ 
is determined by the model-independent relation 
\begin{align}
\label{eq:number_A}
N_A = & K(q_0) N_C(q \le q_0) 
\quad \text{with} \nonumber \\
& K(q_0 = 3.12~\text{MeV}/c) = 0.615 \:,
\end{align}
where $N_C(q \le q_0)$ is the number of Coulomb pairs with $q \le q_0$  
and $K(q_0)$ a known function of $q_0$. 

\begin{figure}[ht]
\begin{center}
\includegraphics[width=0.9\columnwidth]{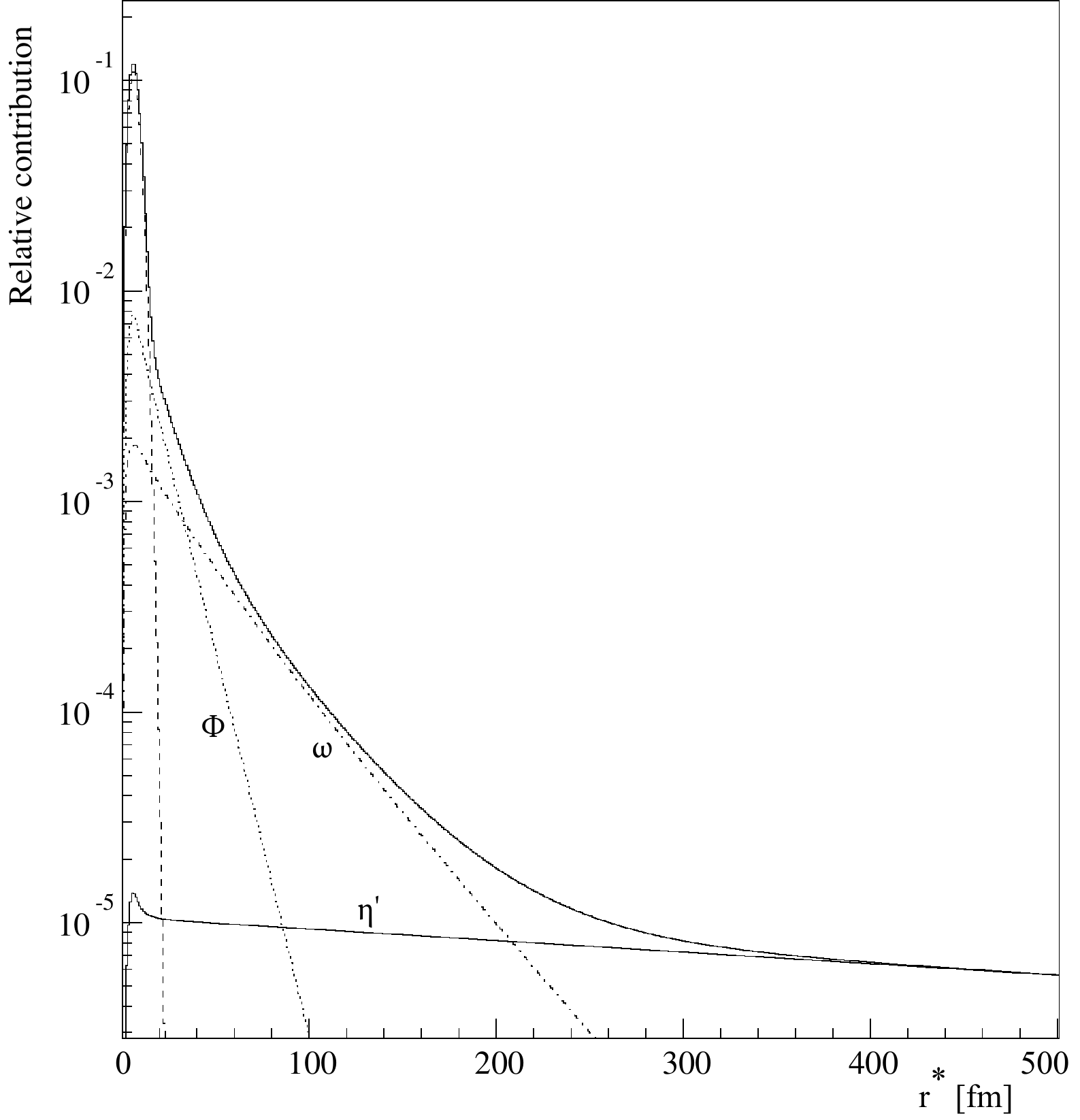}
\caption{Predicted distribution of the relative distance $r^*$ 
between the production points for $\pi K$ pairs. 
The individual curves with increasing $r^*$ correspond to pairs 
produced directly plus from short-lived sources and from  
$\phi$, $\omega$ and $\eta'$ mesons. The sum curve is also shown. }
\label{fig:r_dist}
\end{center}
\end{figure}

Up to now, the pair production was assumed to be pointlike. 
In order to check finite size effects due to the presence 
of medium-lived resonances ($\omega$, $\phi$), a study about 
non-pointlike particle pair sources was performed \cite{LEDN08,note1205}.  
Due to the large value of the Bohr radius $a_B = 249$ fm, 
the pointlike treatment of the Coulomb $\pi K$ FSI is valid 
for directly produced pairs as well as for pairs  
from short-lived strongly decaying resonances. 
This treatment, however, should be adjusted for pions and kaons 
originating from decays of medium-lived particles with path lengths 
comparable with $a_{B}$ in the c.m. system. 
Furthermore, strong FSI should be taken into account: 
elastic $\pi^{+}K^{-} \rightarrow \pi^{+}K^{-}$ or 
$\pi^{-}K^{+} \rightarrow \pi^{-}K^{+}$ 
(driven at $q \rightarrow 0$ by the $S$-wave scattering length 0.137 fm) 
and inelastic scattering $\pi^{0}\bar{K^{0}} \rightarrow \pi^{+}K^{-}$ or 
$\pi^{0}K^{0} \rightarrow \pi^{-}K^{+}$ (scattering length 0.147 fm). 
In Fig.~\ref{fig:r_dist}, the simulated distribution of 
the production regions \cite{LEDN08,note1205} is shown. 
Corrections to the pointlike Coulomb FSI can be performed by 
means of two correction factors $1+\delta(q)$ and $1+\delta_n$ 
($n$ = principal quantum number), to be applied to  
the calculated pointlike production cross sections of Coulomb 
$\pi K$ pairs (\ref{eq:cross_sect_C}) and S-state 
$\pi K$ atoms (\ref{eq:prod}), correspondingly \cite{LEDN08,note1205}. 

\section{Propagation of $\pi K$ atoms through the target}
\label{sec:interact}

To evaluate the $A_{\pi K}$ lifetime from the experimental value of 
the $A_{\pi K}$ breakup probability $P_\mathrm{br}$, it is necessary 
to know $P_\mathrm{br}= f(\tau,l,Z,p_A)$ as a function of 
$A_{\pi K}$ lifetime $\tau$, target thickness $l$, material atomic number $Z$ 
and lab atom momentum $p_A$. After fixing $l$ and $Z$ 
in accordance with the experimental conditions and integrating 
$f(\tau,l,Z,p_A)$ with the measured distribution of $p_A$, 
the dependence $P_\mathrm{br}= f(\tau)$ is obtained. 
\begin{figure}
\begin{center}
\includegraphics[width=\linewidth]{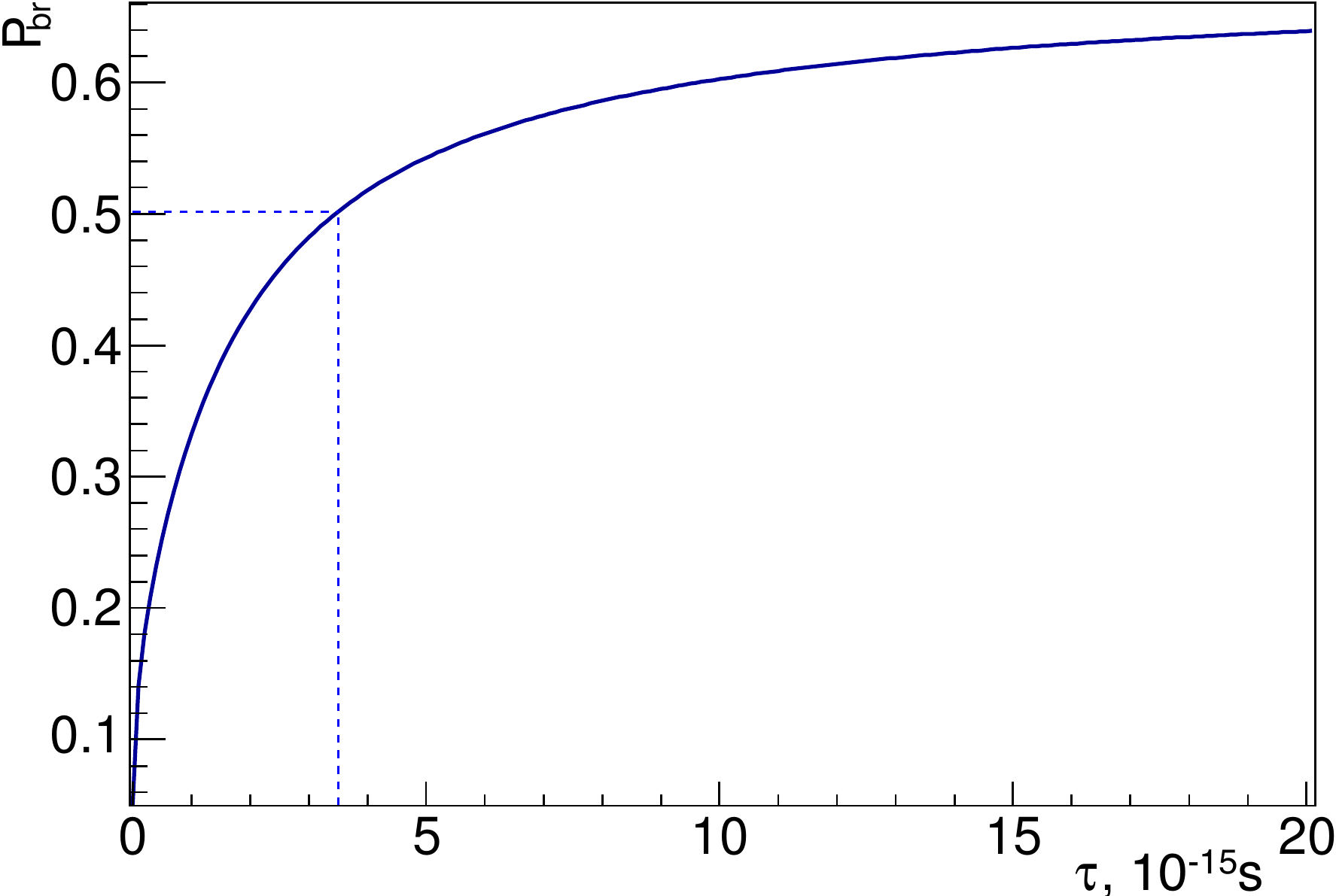}
\includegraphics[width=\linewidth]{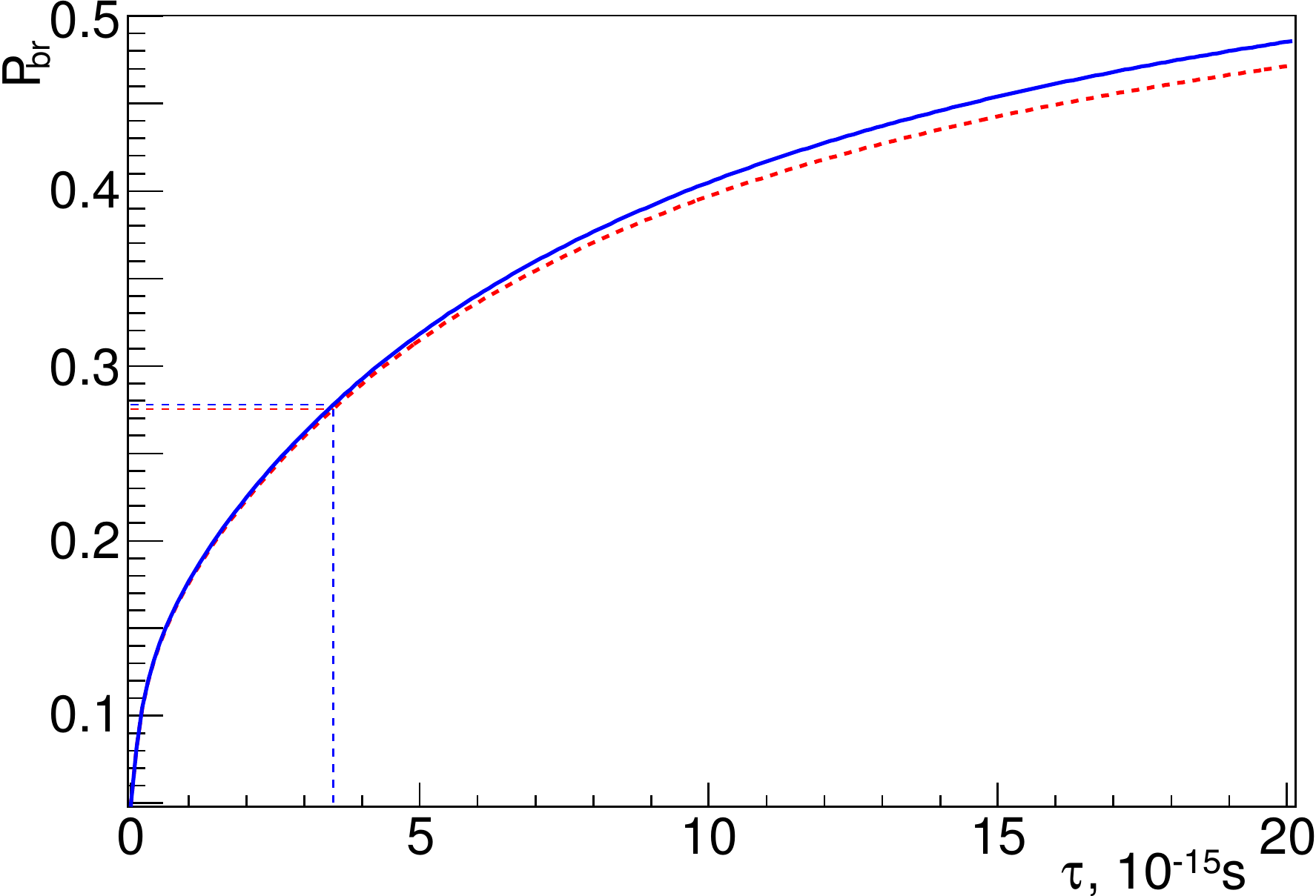}
\end{center}
\caption{
Breakup probability as a function of $\pi K$~atom lifetime $\tau$ 
(ground state) in the DIRAC experiment. 
Top:~Pt~target of thickness 25.7~\textmu{}m. 
Bottom: Ni targets of thicknesses 98~\textmu{}m (red dashed line) and 
108~\textmu{}m (solid blue line).
The predicted lifetime $\tau = 3.5 \cdot 10^{-15}$~s~(\ref{eq:tau35}) 
corresponds to the breakup probabilities $P_\mathrm{br}$ = 0.50~(Pt) 
and 0.28~(Ni).
}
\label{fig:pbr_tau}
\end{figure} 
To calculate $f(\tau,l,Z,p_A)$, one needs to know the total interaction 
cross sections $\sigma_{\mathrm{tot}}(n,l,m)$ of $A_{\pi K}$ with 
matter (ordinary) atoms and all transition (excitation/deexcitation) 
cross sections $\sigma_{if}(n_{i},l_{i},m_{i};n_{f},l_{f},m_{f})$ for 
a large set of initial $i$ and final $f$ $A_{\pi K}$ states 
($n$ principal, $l$ orbital and $m$ magnetic quantum number). 
In the consideration below, all states with $n\le10$ were accounted for. 
Using these cross sections, the distribution of 
the atom quantum numbers at production (\ref{eq:prod}) and 
as free parameter the $A_{\pi K}$ lifetime $\tau$, 
the evolution of each initial $n$S state from the production point 
up to the end of the target is described in order to calculate 
the ionization or breakup probability $P_\mathrm{br}$ (Fig.~\ref{fig:pbr_tau}).

\subsection{Interaction cross sections of $\pi K$ and $\pi \pi$ atoms with
matter atoms}
\label{ssec:cs}

The cross sections of $A_{\pi K}$ interaction with matter atoms were determined
from
analogous theoretical studies about $\pi^+\pi^-$ atoms ($A_{2\pi}$) interacting
with
matter atoms: the $A_{2\pi}$ wave functions are replaced in all formulas by 
the $A_{\pi K}$ wave functions. The interaction of $A_{2\pi}$ with target atoms 
includes two parts: 
1) interaction with screened nuclei, i.e. coherent scattering, that leaves 
the target atom in the initial state and 
2) interaction with orbital electrons, i.e. incoherent scattering, where 
the target atom will be excited or ionized. 
The former is proportional to the square of the nuclear charge ($Z^2$), while 
the latter is proportional to the number of electrons ($Z$). Thus, 
the latter contribution is insignificant for large $Z$. The cross sections 
$\sigma_\mathrm{tot}$ and $\sigma_{if}$ for the coherent interaction are
calculated in first Born approximation (one-photon exchange) by describing the
target atoms in
the Thomas-Fermi model with Moliere parameterisation
\cite{afan96,KOTS80,KOTS83,MROW86,MROW87}.
The $\sigma_{if}$ values taking in to account coherent interaction as well as
the incoherent interactions with more precise non-relativistic Hartree-Fock wave
functions were calculated in \cite{afan96}.
For Ni targets, the incoherent contribution to the cross sections is about 4\% 
of the coherent one.
The influence of relativistic effects on the $\sigma_{if}$ accuracy was studied
\cite{basel00,basel01,basel02}
by describing the ordinary atom with the relativistic Dirac-Hartree-Fock-Slater 
wave functions. Different models for the Ni atom potential lead to 
an uncertainty in $P_\mathrm{br}$ of about 1\% \cite{DN201406}.

In the $A_{2\pi}$ c.m. system, a target atom creates a scalar and a vector
potential.
The interaction with the vector potential (magnetic interaction) was discussed
in \cite{MROW87,basel00,basel01}.
The ``magnetic'' contribution to the cross sections was calculated in
\cite{Afan02}. It was shown that ``magnetic'' contribution to the cross sections
for Ni is about 1\% of the ``electric'' one for $A_{2\pi}$ and
about 2\% for $A_{\pi K}$. All the small cross section corrections discussed
here are about twice larger for $A_{\pi K}$ than for $A_{2\pi}$.

Applying the eikonal (Glauber) approach, the next step in accuracy for the 
mesonium--atom interaction cross sections has been achieved 
\cite{Taras91,Voskr98,basel01,basel02}.
This method includes multi-photon exchange processes in comparison with 
the single-photon exchange in the first Born approximation. 
The total cross sections for the mesonium interaction with ordinary atoms 
were calculated. The interaction cross sections for Ni in this approach 
are less than in the first Born approximation by 0.8\% for $n=1$ and 
at most 1.5\% for $n=6$ \cite{Afan99jp,Ivan99}.
Therefore, the inclusion of multi-photon exchanges is only relevant in
calculations of $\sigma_{if}$ at the 1\% level.
In the above calculations, the target atoms are considered 
isolated, i.e. no solid state modification is applied to the wave functions. 
A dedicated analysis \cite{basel01} proves that solid-state effects and 
target chemistry do not change the $A_{2\pi}$ cross sections.
In the mentioned cross section calculations, the $A_{2\pi}$ wave functions
are the hydrogen-like non-relativistic Schr\"odinger equation solutions. 
The relativistic Klein--Gordon equation for the $A_{2\pi}$ description leads 
to negligible relativistic corrections to the cross sections \cite{basel00}. 
Furthermore, the seagull diagram contribution can be safely 
neglected \cite{hadatom01}. 

\subsection{$\pi K$ and $\pi \pi$ atom breakup probabilities}
\label{ssec:br}

The description of the $A_{\pi K}$ (multilevel atomic system) propagation 
in (target) matter is almost the same as in the case for $A_{2\pi}$, 
first considered in \cite{afan96}. $A_{2\pi}$, produced in 
proton-nucleus collisions, can either annihilate or interact with target atoms. 
It was shown that stationary atomic states are formed between 
two successive interactions, at least for $n \leq 6$. 
Thus, the population of each level can be described in terms of probabilities, 
disregarding interferences between degenerated states with the same energy. 
The population of atomic $A_{2\pi}$ states, moving in the target, is 
described by a set of differential (kinetic) equations, accounting  
for the $A_{2\pi}$ interaction with target atoms and 
the $A_{2\pi}$ annihilation. The set of kinetic equations, 
formally containing an infinite number of equations, is truncated up to 
states with $n\leq7$ to get a numerical solution. 
The breakup probability $P_\mathrm{br}$ is calculated by applying 
the unitary condition:
\[
P_\mathrm{br}+ P_\mathrm{dsc}(n \leq 7) + P_\mathrm{dsc}(n>7) + P_\mathrm{ann} 
= 1,
\]
where $P_\mathrm{dsc}(n \leq 7)$ and $P_\mathrm{dsc}(n>7)$ are 
the populations of the discrete $A_{2\pi}$ states, leaving the target, 
with $n \leq 7$ and $n>7$, and $P_\mathrm{ann}$ is 
the $A_{2\pi}$ annihilation probability in the target. Values of 
$P_\mathrm{dsc}(n \leq 7)$ and $P_\mathrm{ann}$ are obtained by solving 
the truncated set of kinetic equations. On the other hand, one gets 
a value of $P_\mathrm{dsc}(n>7)$ by extrapolating the calculated behaviour of 
$P_\mathrm{dsc}(n)$. The value of $P_\mathrm{dsc}(n>7)$ is about 0.006, 
and the extrapolation accuracy is insignificant for the accuracy of
$P_\mathrm{br}$. The method here only uses total cross sections
and transition cross sections between discrete $A_{2\pi}$ states. 

Obtaining the ionization (breakup) cross sections for 
an arbitrary $A_{2\pi}$ bound state \cite{basel99,basel00}, allows 
to calculate directly $P_\mathrm{br}$ \cite{Zhab08}. The difference of 0.5\% 
between two methods for $n=8$ demonstrates the convergence and estimates 
the $P_\mathrm{br}$ precision.

To clarify the influence of the interference between degenerated states 
with the same energy, the motion of $A_{2\pi}$ in 
the target was described in the density matrix formalism \cite{Voskr03}.  
The $P_\mathrm{br}$ value calculated using this method coincides with the one in
the probability based approach with an accuracy of better than $10^{-5}$
\cite{Afan04}. The same is true for $A_{\pi K}$.

The function $P_\mathrm{br}=f(\tau,l,Z,p_A)$ has a weak dependence on 
the target thickness $l$ in the conditions of the DIRAC experiment. 
The relative $l$ uncertainty of $\pm$1\% leads to an insignificant error 
of $f(\tau,l,Z,p_A)$ on the level of $\pm$0.1\%. 

In the present article, $P_\mathrm{br}=f(\tau,l,Z,p_A)$ is  
calculated by means of the DIPGEN code \cite{DIPGEN}, using the unitary
condition
and the set of $A_{\pi K}$ total and transition cross sections calculated in 
the approach of Ref.~\cite{afan96} for $n\leq10$ without 
taking into account the incoherent interaction, magnetic interaction and 
multi-photon exchange \cite{pik-prop}. As described above, all these effects 
contribute to the cross section only at the level of (1--2)\% with 
different signs. The common error of the approximation used is evaluated in 
the following way. The $A_{2\pi}$ breakup probabilities 
$P_\mathrm{br}^{\pi\pi}$ are determined in the same way as for $A_{\pi K}$ and 
also using very precise cross sections \cite{basel99,basel00,basel01,basel02} 
considering all types of interactions. The difference in 
the $P_\mathrm{br}^{\pi\pi}$ values is 0.6\% \cite{pik-prop}. 
For $A_{\pi K}$, the contributions of unaccounted cross sections are larger 
than for $A_{2\pi}$ (see above). Hence, the difference in $P_\mathrm{br}$ 
is expected to be larger by a factor of around 2. The accuracy of 
the $P_\mathrm{br}$ calculation procedure for Ni is estimated as 
0.8\% \cite{Zhab08}. Therefore, the upper limit of the total uncertainty of 
$P_\mathrm{br}$ for $A_{\pi K}$ cannot exceed 2\%, compared to 1\% for 
$A_{2\pi}$ \cite{ADEV11}. This value is significantly smaller than 
the statistical accuracy.

\subsection{Relative momentum distribution of atomic $\pi K$ pairs}
\label{ssec:q}

The evaluation of the number of the atomic pairs requires the knowledge 
of their distribution on the relative momentum at the target exit and after 
the reconstruction. This distribution depends on the atomic quantum numbers 
at the atom breakup point and the coordinates of this point. 
The relative momentum distributions of the atomic pairs for different 
atom quantum numbers have been calculated \cite{hadatom01} and were entered 
into DIPGEN \cite{DIPGEN}. This distribution is further broadened by 
multiple scattering of the mesons in the target. 
The main influence on the distribution of 
the transverse relative atomic pair momentum at the target exit 
is due to multiple scattering in the target, whereas the influence 
from the atomic states is significantly smaller, but nevertheless taken 
into account in DIPGEN.

\section{Data processing} 
\label{sec:dp}

The collected events were analyzed with the DIRAC reconstruction program 
ARIANE \cite{Ariane} modified for analyzing $\pi K$ data. 

\subsection{Tracking} 
\label{ssec:track}

Only events with one or two particle tracks in DC   
of each arm are processed. The event reconstruction is 
performed according to the following steps:
\begin{itemize}
\item One or two hadron tracks are identified in DC  
of each arm with hits in VH, HH and PSh slabs and 
no signal in ChN and Mu.
\item Track segments, reconstructed in DC, are extrapolated 
backward to the beam position in the target, 
using the transfer function of the dipole magnet and 
the program ARIANE. This procedure 
provides approximate particle momenta and  
the corresponding points of intersection in MDC, SFD and IH.
\item Hits are searched for around the expected SFD coordinates 
in the region $\pm 1$~cm corresponding to (3--5)~$\sigma_{\mathrm{pos}}$ 
defined by the position accuracy taking into account 
the particle momenta. The number of hits around the two tracks is 
$\le 4$ in each SFD plane and  $\le 9$ in all three SFD planes. 
The case of only one hit in the region $\pm 1$~cm can occur 
because of detector inefficiency (two crossing particles, but  
one is not detected) or if two particles cross the same SFD column. 
The latter type of event may be recovered by selecting double ionization   
in the corresponding IH slab. 
For RUN1 data collected with the Pt target, the criteria are different: 
the number of hits is two in the $Y$- and $U$-plane (signals from 
SFD $X$-plane and IH, which may resolve crossing of 
only one SFD column by two particles, were not available in RUN1 data). 
\end{itemize}
The momentum of the positively or negatively charged particle 
is refined to match the $X$-coordinates of the DC tracks 
as well as the SFD hits in the $X$- or $U$-plane, 
depending on the presence of hits. In order to find 
the best 2-track combination, the two tracks may not use 
a common SFD hit in the case of more than one hit in the proper region. 
In the final analysis, the combination with the best $\chi^2$ 
in the other SFD planes is kept. 

\subsection{Setup tuning using $\Lambda$ and $\bar{\Lambda}$ particles} 
\label{ssec:geom}

In order to check the general geometry of the DIRAC experiment, 
the $\Lambda$ and $\bar{\Lambda}$ particles, decaying into 
${\rm p}\pi^-$ and $\pi^+\bar{\rm p}$ in our setup, were used. 
Details of this study are reported in \cite{DN201601,lan,note0516}. 
Comparing our reconstructed $\Lambda$ mass values with PDG data \cite{pdg} 
allows to check the geometrical setup description. The main factors, 
that can influence the value of the $\Lambda$ mass, are 
the position of the aluminium (Al) membrane (defining the location of 
the spectrometer magnetic field relative to the setup detectors) and 
the angles between each downstream telescope arm axis 
and the setup axis (secondary particle beam direction). 
The position of the Al membrane was fixed to $z_{Al} = 1433.85$~mm from 
the centre of the magnet. The orientation of the downstream arm axes 
should be corrected on average for the right arm by $-0.032$ mrad 
and for the left arm by $+0.088$ mrad relative to the geodesic measurements. 
The values, from year to year used, are reported in \cite{DN201601}. 

\begin{figure}[h]
\centering 
\includegraphics[width=0.8\columnwidth]{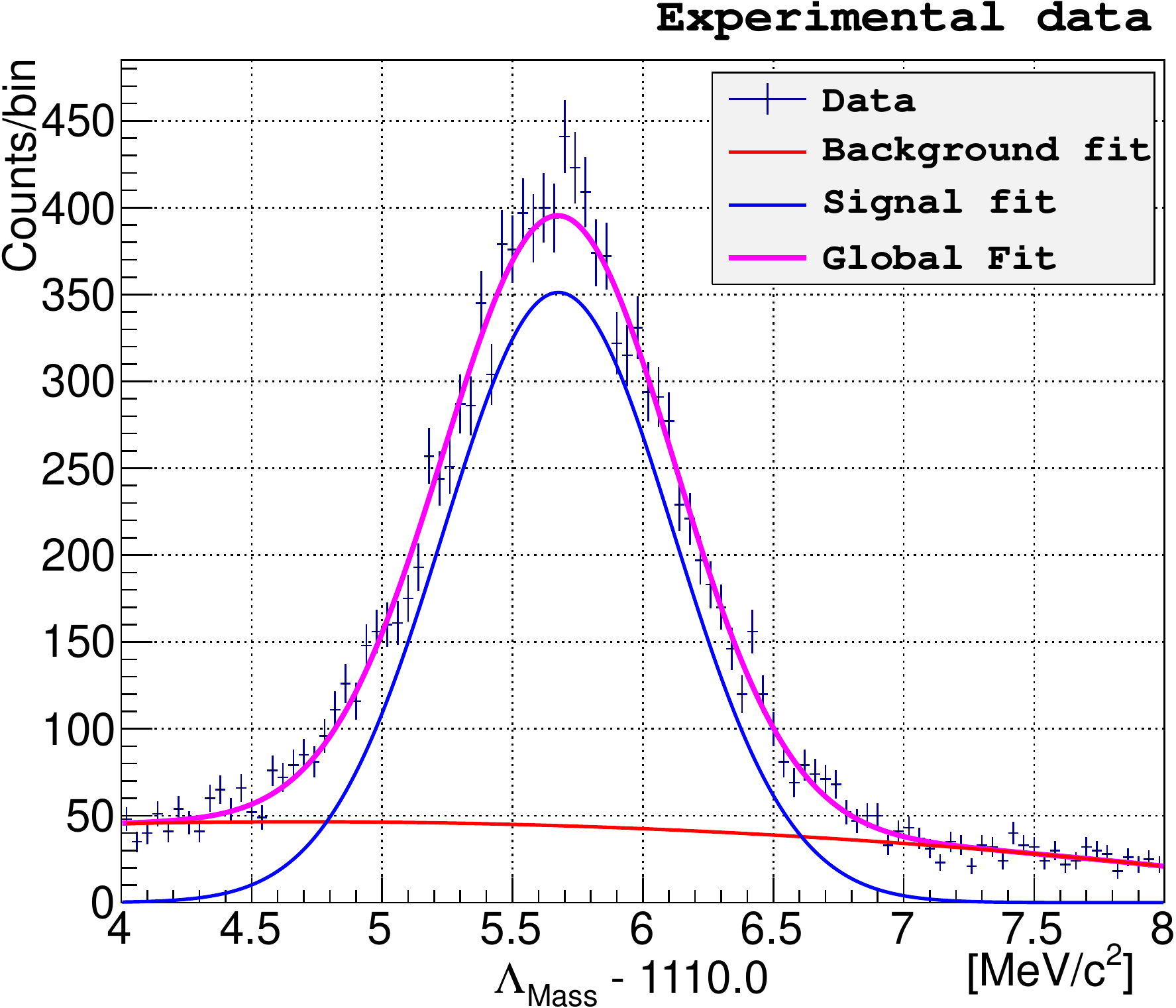}
\includegraphics[width=0.8\columnwidth]{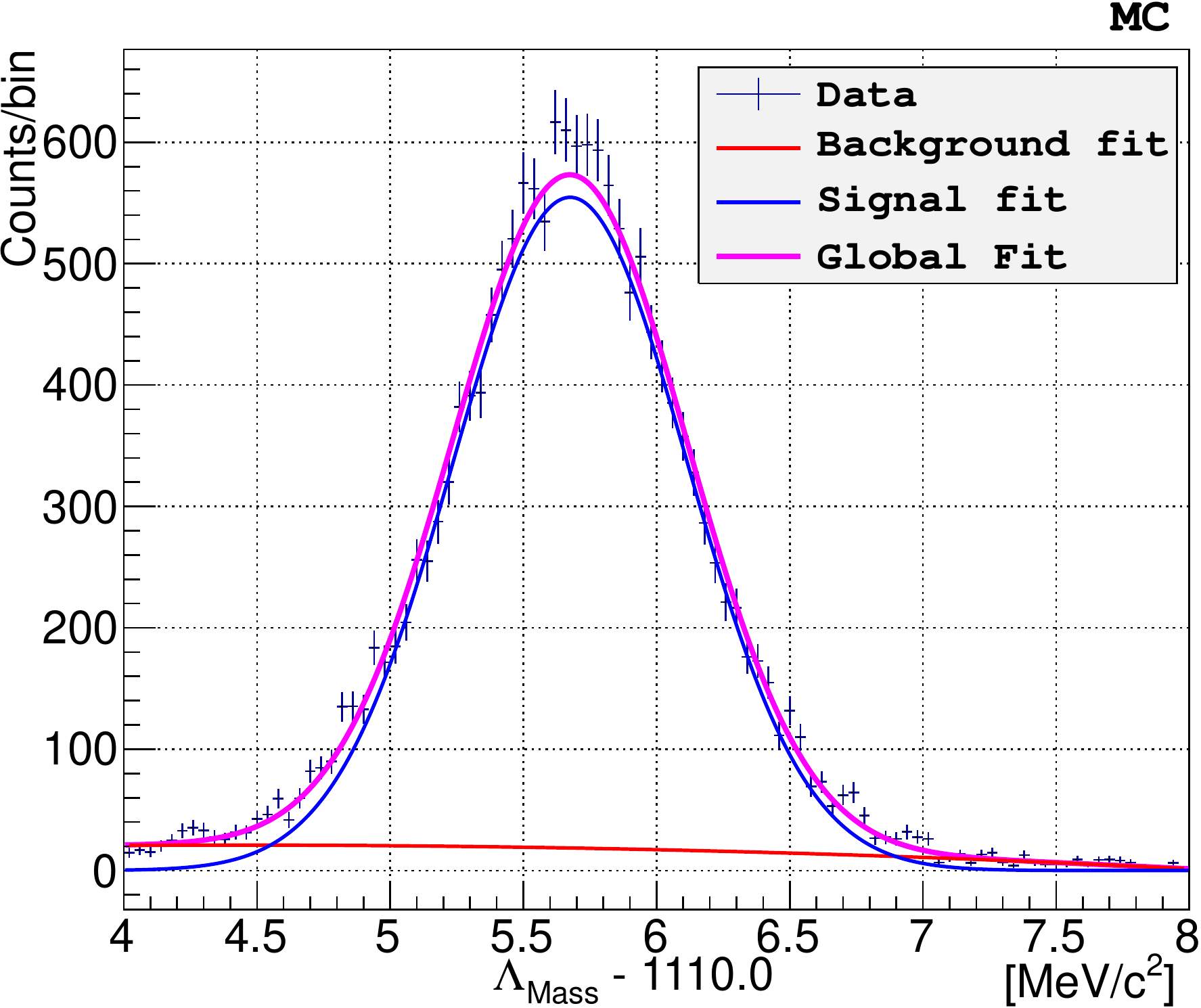}
\caption{$\Lambda$ mass distribution for RUN3 data (top) and 
MC simulation (bottom) are fitted with a Gaussian (in blue) for
the $\Lambda$ peak and a second degree polynomial (in red) describing the 
background. $\Lambda^{exp} -1110.0 =5.676  \pm 5.9 \cdot 10^{-3}$ 
and $\Lambda^{MC} -1110.0 =5.675 \pm 4.3 \cdot 10^{-3}$ in $\rm{MeV}/c^2$.}
\label{fig:lambda}
\end{figure}

Fig.~\ref{fig:lambda} shows the distribution of the $\Lambda$ mass for the RUN3
data
and for the corresponding Monte Carlo (MC) simulation. The distributions are
fitted
with a Gaussian and a second degree polynomial that describes the background. 
The weighted average value of the experimental $\Lambda$ mass over all runs, 
$M^\mathrm{DIRAC}_{\Lambda} = (1.115680 \pm 2.9 \cdot 10^{-6}) \;
\text{GeV}/c^2$,
agrees very well with the PDG value,  
$M^{\text{PDG}}_{\Lambda} = (1.115683 \pm 6 \cdot 10^{-6}) \; \text{GeV}/c^2$. 
The weighted average of the experimental ${\bar{\Lambda}}$ mass is 
$M^\mathrm{DIRAC}_{\bar{\Lambda}} = (1.11566 \pm 1 \cdot 10^{-5}) \;
\mathrm{GeV}/c^2$.
This demonstrates that the geometry of the DIRAC setup is well described. 

The width of the $\Lambda$ mass distribution allows to test the momentum and 
angular setup resolution in the simulation. Table ~\ref{tab:lambda-w} shows 
a good agreement between simulated and experimental $\Lambda$ width. 
A further test consists in comparing the experimental $\Lambda$ and 
${\bar{\Lambda}}$ widths.

\begin{table}[htbp]
\caption{Standard deviations from Gaussian fit of $\Lambda$ peak in GeV/$c^2$
for experimental and MC data and ${\bar\Lambda}$ experimental data.}
\label{tab:lambda-w}
\begin{ruledtabular}
\begin{tabular}{cccc}
& $\sigma_\Lambda$ (data) & $\sigma_\Lambda$ (MC) & $\sigma_{\bar{\Lambda}}$
(data) \\
&      $\rm{GeV}/c^2$     &     $\rm{GeV}/c^2$   &     $\rm{GeV}/c^2$ \\ \hline
RUN1 &    $4.22 \cdot 10^{-4}$ & $ 4.15 \cdot 10^{-4}$ & $4.3 \cdot 10^{-4}$ \\
&${}\pm 4.6\cdot 10^{-6}$ & $ {}\pm 2.9\cdot 10^{-6}$ & ${}\pm 3\cdot 10^{-5}$
\\ \hline
RUN2 & $4.33 \cdot 10^{-4}$ & $ 4.38 \cdot 10^{-4}$ & $4.6 \cdot 10^{-4}$ \\
&${}\pm 8.2 \cdot 10^{-6}$ & $ {}\pm 4.6 \cdot 10^{-6}$ & 
${}\pm 2 \cdot 10^{-5}$ \\ \hline
RUN3 & $4.42 \cdot 10^{-4}$ & $ 4.42 \cdot 10^{-4}$ & $4.5 \cdot 10^{-4}$ \\
& ${}\pm 7.4 \cdot 10^{-6}$ & $ {}\pm 4.4 \cdot 10^{-6}$ & 
${}\pm 3 \cdot 10^{-5}$ \\ \hline
RUN4 &    $4.41 \cdot 10^{-4}$ & $ 4.37 \cdot 10^{-4}$ & $4.3\cdot 10^{-4}$ \\
& ${}\pm 7.5\cdot 10^{-6}$ & $ {}\pm 4.5\cdot 10^{-6}$ & ${}\pm 2\cdot 10^{-5}$
\\
\end{tabular}
\end{ruledtabular}
\end{table}

In order to understand, if the differences between data and MC are 
significant or just due to statistical fluctuations, the MC distributions 
were generated with a width artificially squeezed and enlarged.
In every simulated event, the value of the reconstructed invariant mass
of the system pion-proton, $x$, was modified according to 
$MC_f = ( x - M_{MC} ) \cdot f + M_{DATA}$, where $f$ is the parameter 
shrinking or enlarging the $\Lambda$ distribution by $\pm 20\%$ in steps of
2\%.
The $\Lambda$ peak positions of the experimental and original MC distributions 
are denoted by $M_\mathrm{DATA}$ and $M_\mathrm{MC}$, respectively. Then, the
experimental
and modified MC distributions were compared \cite{CERN-EP-2017-137}. For RUN1
with the Pt target and 2 SFD planes, procedure found the best agreement for
$f_{RUN1}=1.019 \pm 2. \cdot 10^{-3}$. For the runs with 3 SFD planes and Ni
target,
the following $f$ values were obtained:
$f_{RUN2} = 1.00235 \pm 4.34 \cdot 10^{-3}$,
$f_{RUN3} = 1.00059 \pm 2.75 \cdot 10^{-3}$ and
$f_{RUN4} = 1.00401 \pm 3.38~\cdot~10^{-3}$
with the average value $f_{Ni} = 1.00203 \pm 0.00191$.

The difference between data and MC widths could be the consequence of
imperfectly describing the downstream setup part, to be fixed by
a Gaussian smearing of the reconstructed momenta for MC data. On an
event--by--event basis, the smearing of the reconstructed proton and pion
momentum $p$ has been applied in the form
$p^\mathrm{smeared} = p(1+C \cdot N(10^{-4}))$,
where $N(10^{-4})$ is a normally distributed random number with a mean of 0 and
a standard deviation of 0.0001. The values $f_{RUN1}$ and $f_{Ni}$ correspond to
$C_{RUN1}=6.7 ^{+2.2}_{-2.9}$ and $C_{Ni}=2.2319^{+0.7438}_{-1.1758}$,
respectively.

\begin{figure}[h]
\begin{center}
\includegraphics[width=\linewidth]{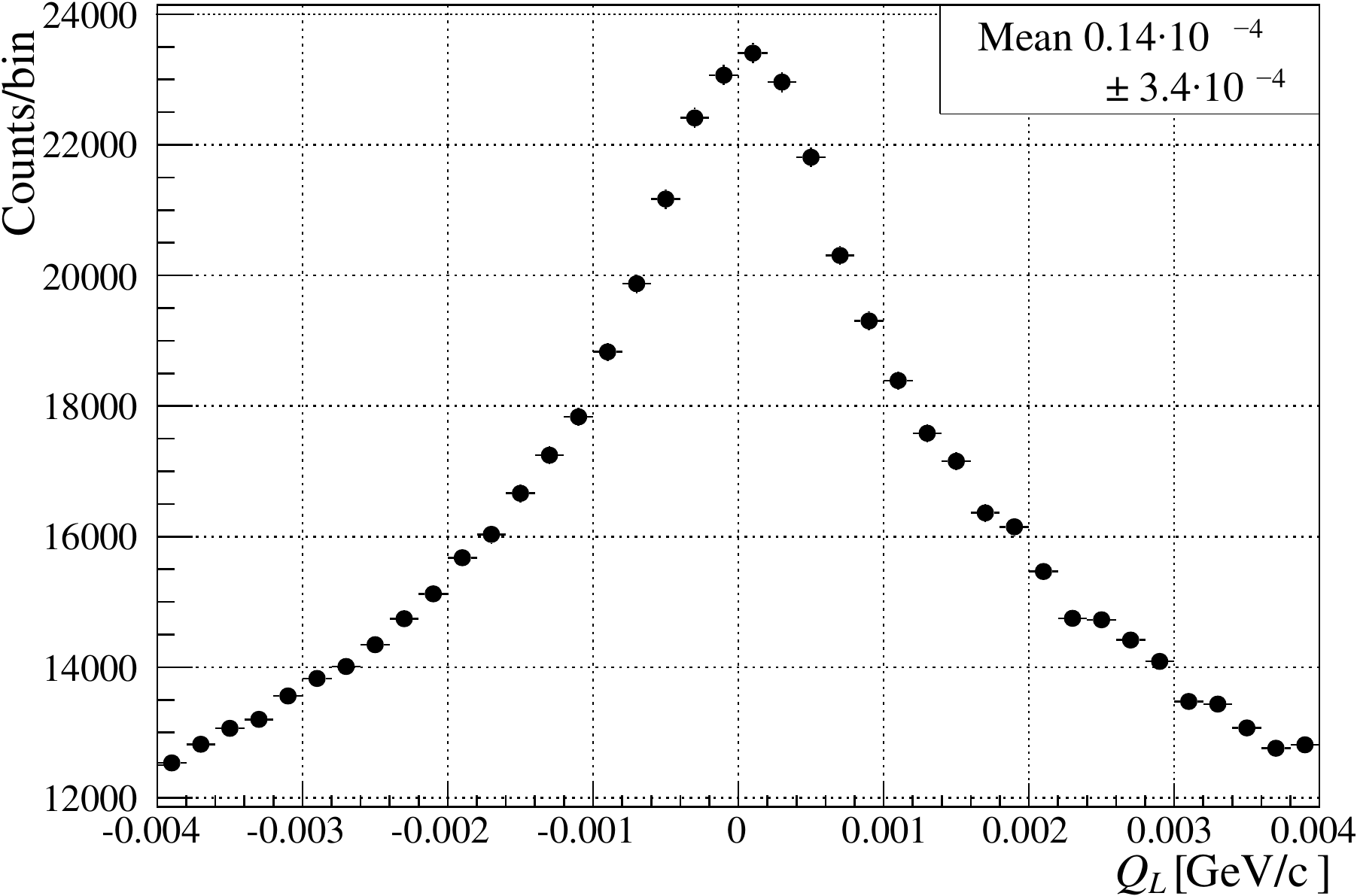}
\end{center}
\caption{$Q_L$ distribution of $\pi^+\pi^-$ experimental data with cut 
	$Q_T < 4~\text{MeV}/c$ (RUN2 to RUN4). }
\label{fig:6_3}
\end{figure}

The $Q_L$ distribution of 
$\pi^+\pi^-$ pairs can be used to check the geometrical alignment. 
Since the $\pi^+\pi^-$ system is symmetric, 
the corresponding $Q_L$ distribution should be centered at 0. 
Fig.~\ref{fig:6_3} shows the experimental $Q_L$ distribution of 
pion pairs with transverse momenta $Q_T < 4~\text{MeV}/c$: 
the distribution is centered at 0 with a precision of 0.2~$\text{MeV}/c$.

\subsection{Background subtraction} 
\label{sec:backr_subtr}

The background of electron-positron pairs is suppressed by ChN at 
the first level of the trigger system. Because of the large 
$\text{e}^+\text{e}^-$ flux and finite ChN efficiency, 
a certain admixture of $\text{e}^+\text{e}^-$ pairs with 
small $Q_T$ remains and can induce a bias in the data analysis. 
To further suppress this background, the preshower scintillation detector 
PSh is used~\cite{PENT15}. 

\begin{figure}[h]
\begin{center}
\includegraphics[width=\columnwidth]{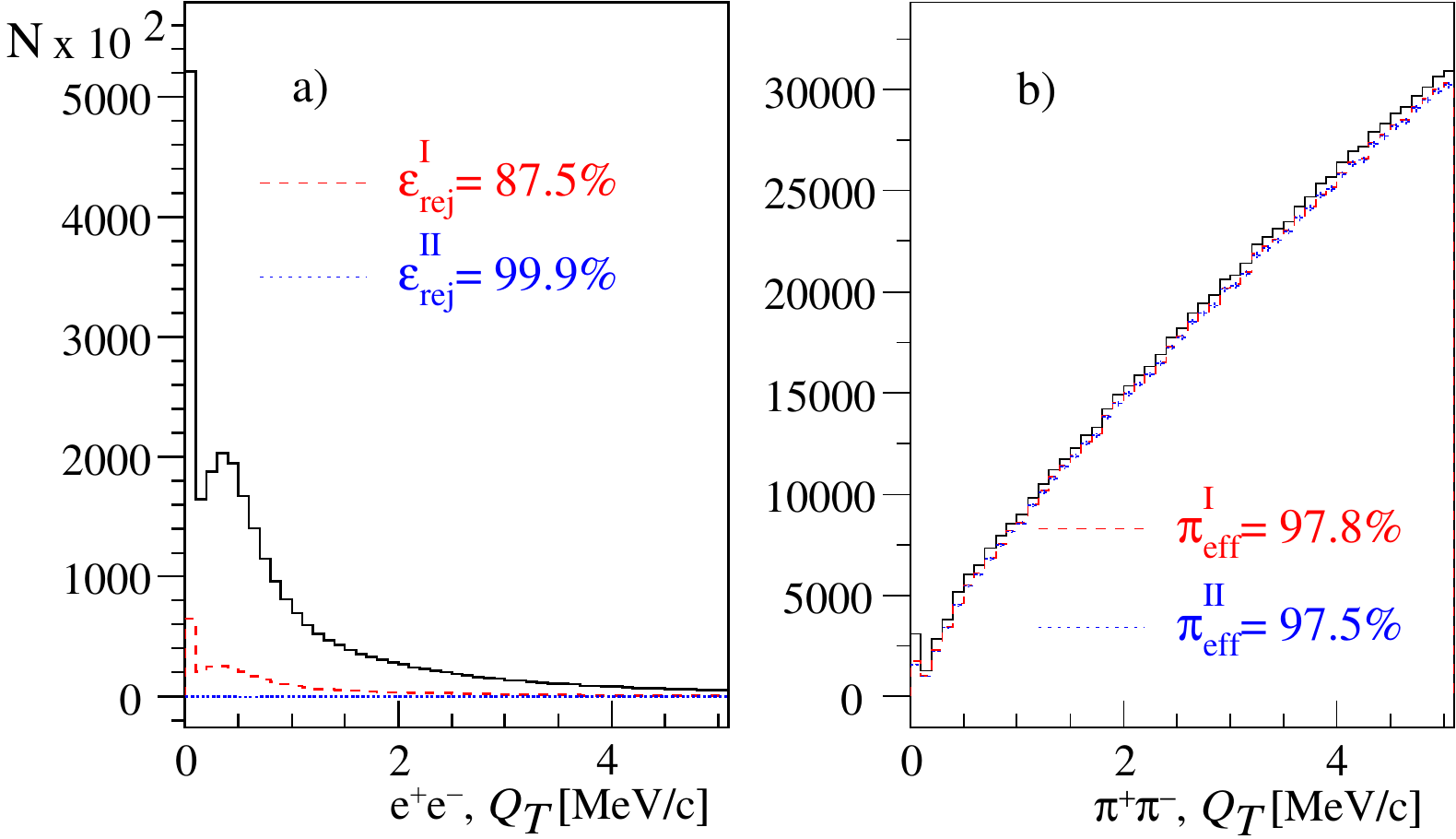}
\end{center}
\caption{$Q_T$ distributions for a) $\text{e}^+\text{e}^-$ and b) $\pi^+\pi^-$
pairs without PSh amplitude criterion (black solid line),  
after amplitude criterion (red dashed line) and after additional subtraction 
of electron admixture in the accepted events (blue dotted line).}
\label{fig:qt_ee_pipi}
\end{figure}

At the preparation stage, a set of $\pi^+\pi^-$ (hadron-hadron) and 
a set of $\text{e}^+\text{e}^-$ data were selected by using ChN 
(low and high amplitude in both arms, respectively). For each pair of 
PSh slabs ($i$-th slab in the left and $j$-th in the right arm), a procedure 
selects the amplitude criterion of these slabs accepting 98\% of 
the $\pi^+\pi^-$ and suppressing $\text{e}^+\text{e}^-$ pairs. Furthermore, 
the ratio $R_{ij}$ of $\text{e}^+\text{e}^-$ events accepted 
($N^{\text{accepted}}_{ij}$) and rejected 
($N^{\text{rejected}}_{ij}$) by this criterion was calculated for 
electron trigger data: 
$ R_{ij}=\frac{N^{\text{accepted}}_{ij}}{N^{\text{rejected}}_{ij}} $. 
In the data analysis, these criteria are applied to the events. 
Fig.~\ref{fig:qt_ee_pipi}a and Fig.~\ref{fig:qt_ee_pipi}b present 
the results for $\text{e}^+\text{e}^-$ pairs and $\pi^+\pi^-$ pairs,
respectively.
The initial distributions are shown as black solid lines and 
the distributions after applying the PSh amplitude criterion 
in the left and right arm as red dashed lines. 
This criterion accepts 97.8\% of $\pi^+\pi^-$ pairs and rejects 87.5\% of   
$\text{e}^+\text{e}^-$ pairs. To improve the $\text{e}^+\text{e}^-$ suppression, 
the remaining electron admixture in the PSh cut data is subtracted from 
the distribution of accepted events with the event-by-event weight 
$R_{ij}$. The final distributions are shown as 
blue dotted lines. The rejection efficiency for the $\text{e}^+\text{e}^-$
background
achieves 99.9\%, whereas 2.5\% of the $\pi^+\pi^-$ data are lost.

\begin{figure}[ht]
\begin{center}
\includegraphics[width=0.9\columnwidth,height=0.8\columnwidth]{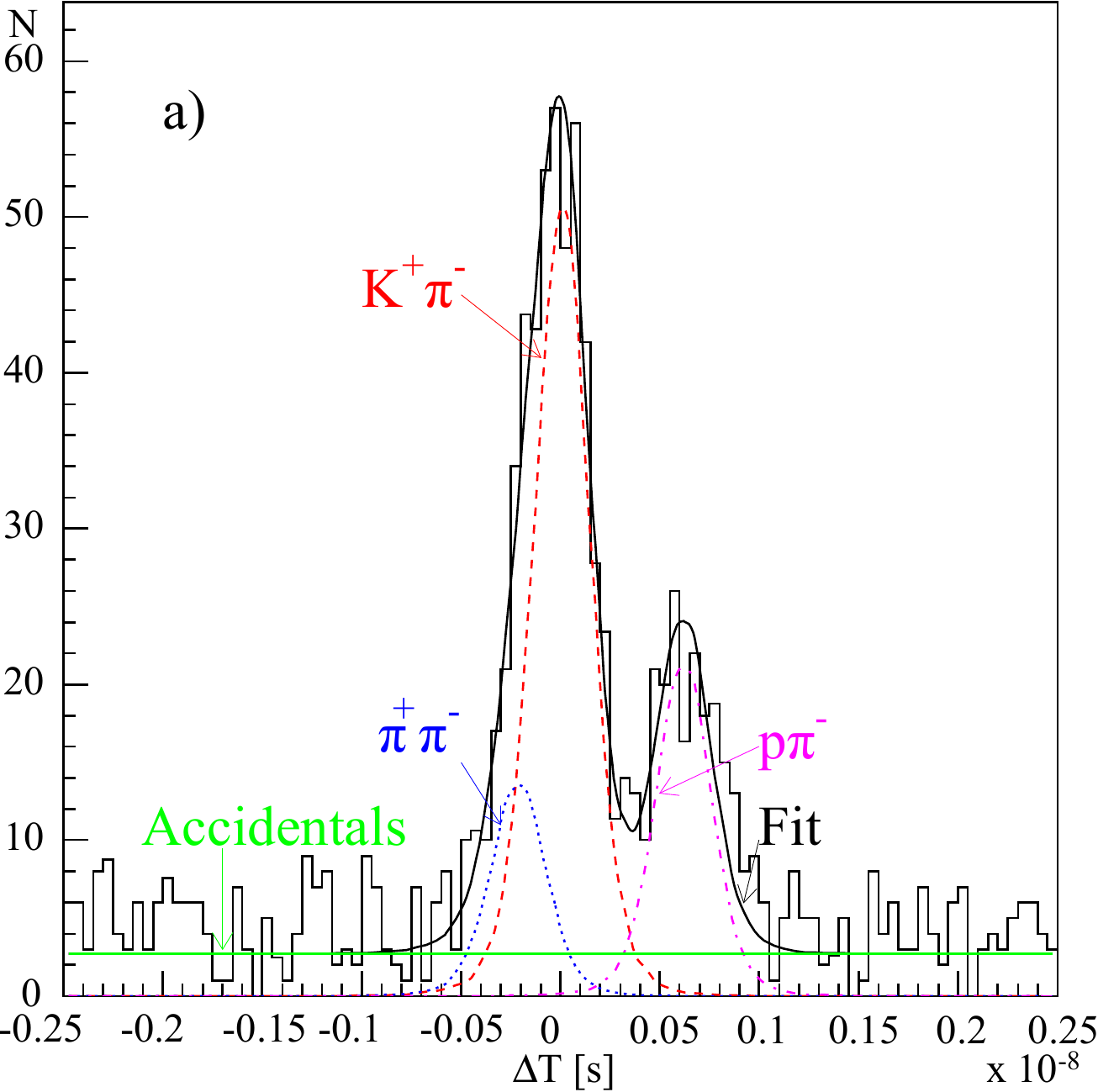}
\includegraphics[width=0.9\columnwidth,height=0.8\columnwidth]{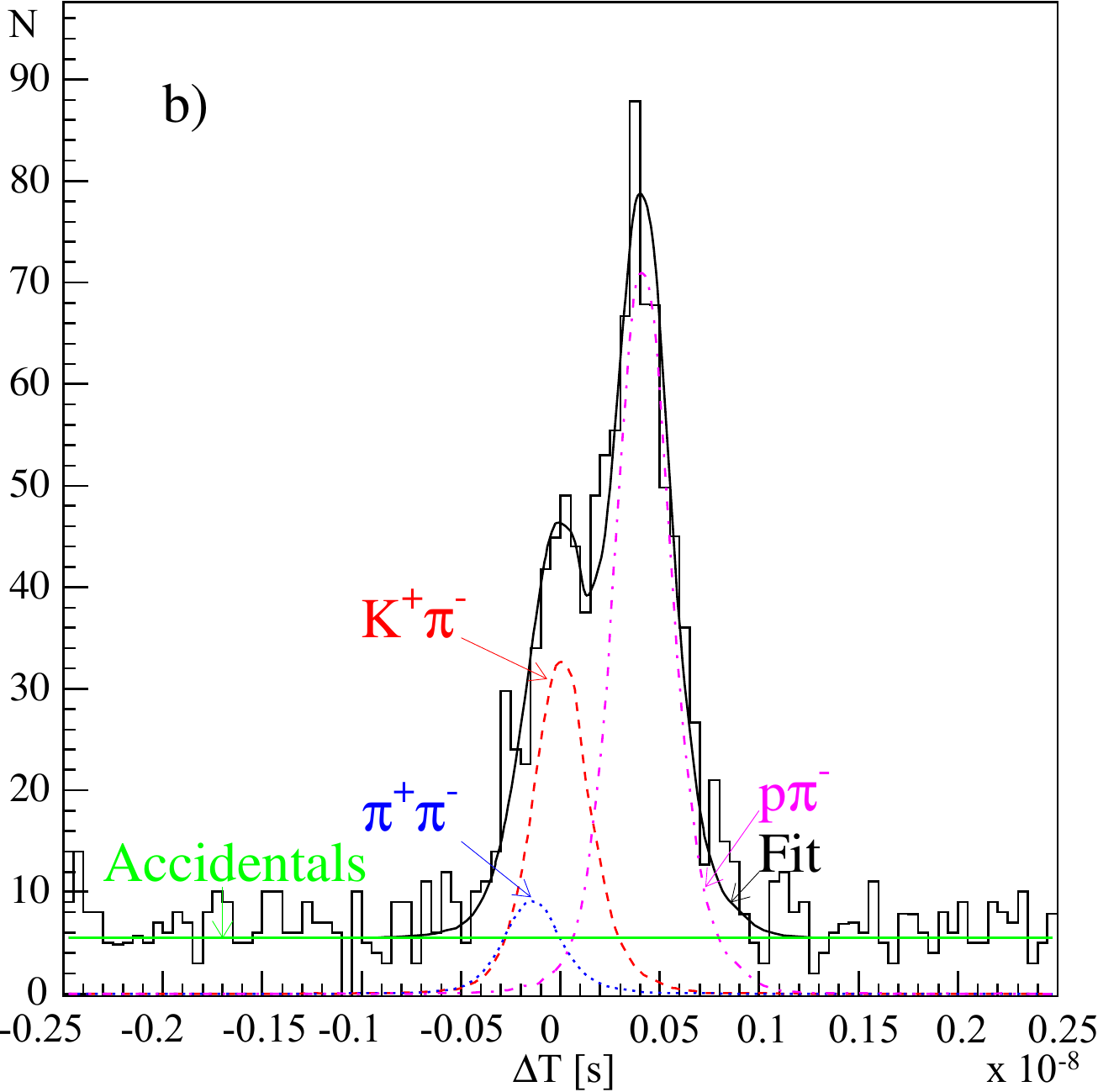}
\end{center}
\caption{a) Difference of particle generation times for events with 
positively charged particle momenta $(4.4 \div 4.5)~\text{GeV}/c$. 
Experimental data (histogram) are fitted by the event sum 
(black, solid): $K^+\pi^-$ (red, dashed), $\pi^+\pi^-$ 
(blue, dotted), $\text{p}\pi^-$ (magenta, dotted-dashed) and 
accidentals (green, constant). b) Similar distributions for events with 
positively charged particle momenta $(5.4 \div 5.5)~\text{GeV}/c$.}
\label{fig:dTmGen}
\end{figure}

\subsection{Event selection criteria} 
\label{ssec:Ev_sel}

\begin{figure*}[]
\begin{center}
\includegraphics[width=\linewidth]{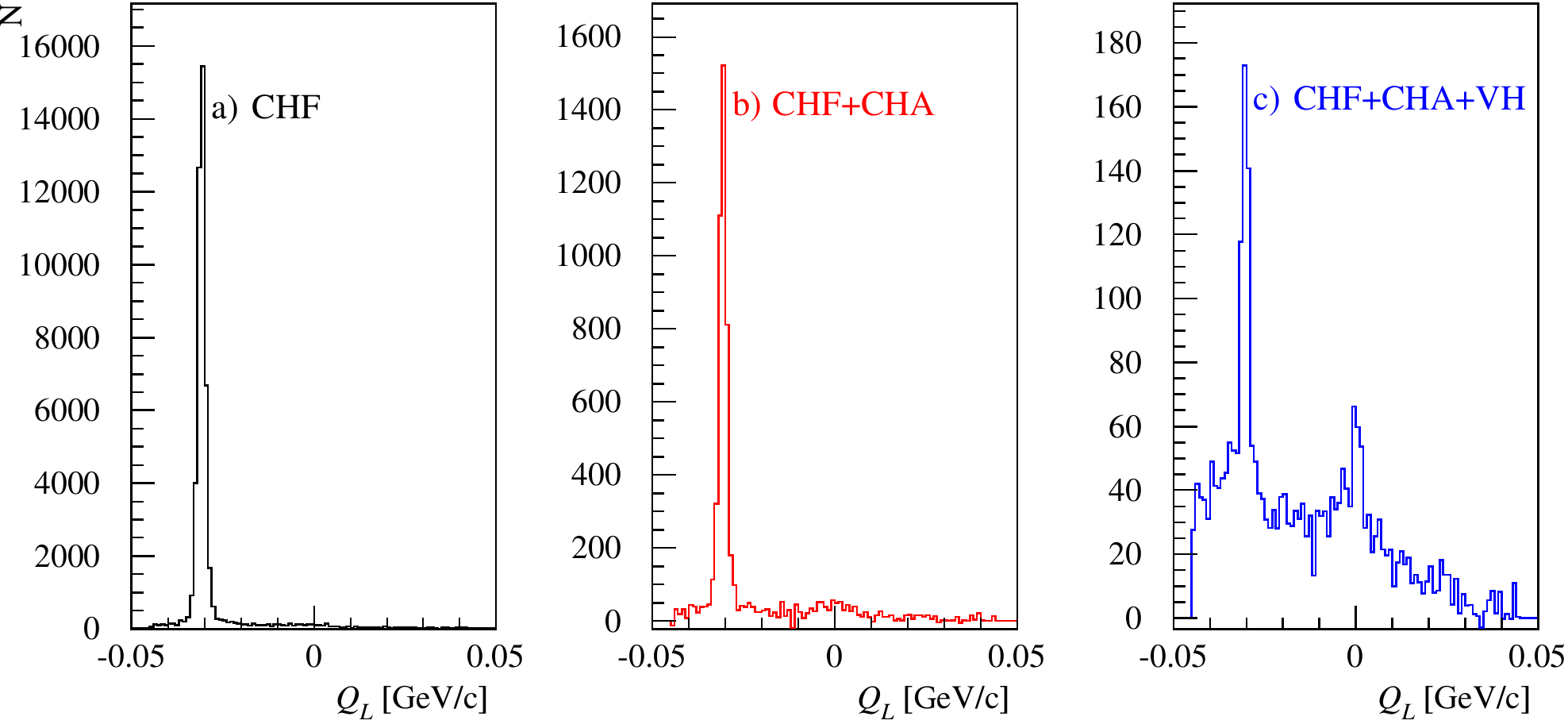}
\caption{$Q_L$ distribution of $\pi^-K^+$ pairs after applying 
different criteria (see text).}
\label{fig:kpi-selection}
\end{center}
\end{figure*}

The selected events are classified into three categories: 
$\pi^-K^+$, $\pi^+K^-$ and $\pi^-\pi^+$. 
The last category is used for calibration. 
Pairs of $\pi K$ are cleaned of $\pi^-\pi^+$ and $\pi^-{\rm p}$ 
background by the Cherenkov counters ChF and ChA 
(Section~\ref{sec:setup}). In the momentum range from 3.8 to 7~$\text{GeV}/c$, 
pions are detected by ChF with (95--97)\% efficiency~\cite{note1305}, 
whereas kaons and protons (antiprotons) do not produce any signal. 
The admixture of $\pi^-{\rm p}$ pairs is suppressed by ChA, 
which records kaons but not protons \cite{note0907}. 
Due to finite detector efficiency, a certain admixture of misidentified
pairs still remains in the experimental distributions. 
For the selected events, the procedure applied plots the distribution 
of the measured difference $\Delta T$ of particle generation times. 
These times of production at the target are the times, 
which are measured by VH and reduced by the time-of-flights from 
the target to the VH planes for particles with the expected masses 
($K^{\pm}$ and $\pi^{\mp}$ mesons) and the measured lab momenta.
For $\pi^-K^+$ ($\pi^+K^-$) pairs, the difference is centered at 0 
and, for misidentified pairs, biased. 
Fig.~\ref{fig:dTmGen}a presents the event distribution over 
the difference of the particle production times for $K^+$ mesons in 
the range (4.4--4.5)~$\text{GeV}/c$. The distribution is fitted by the 
simulated distribution of admixed fractions. Similarly to 
Fig.~\ref{fig:dTmGen}a, Fig.~\ref{fig:dTmGen}b shows the fit for $K^+$ in 
the range (5.4--5.5)~$\text{GeV}/c$.
The contribution of misidentified pairs was estimated and accordingly 
subtracted \cite{note1306}. Fig.~\ref{fig:kpi-selection}a illustrates 
the $Q_L$ distribution of potential $\pi^-K^+$ pairs requiring 
a ChF signal and $Q_T < 4~\text{MeV}/c$. The dominant peak on the left side 
is due to ${\rm p}\pi^-$ pairs from $\Lambda$ decay. 
After requesting a ChA signal, the admixture of ${\rm p}\pi^-$ pairs is
decreased by a factor of 10 (Fig.~\ref{fig:kpi-selection}b). By selecting 
compatible TOFs between target and VH, background $\text{p}\pi^-$ and
$\pi^+\pi^-$
pairs can be substantially suppressed (Fig.~\ref{fig:kpi-selection}c). 
In the final distribution, the well-defined $\pi^-K^+$ Coulomb peak 
at $Q_L=0$ emerges beside the strongly reduced peak from $\Lambda$ decays 
at $Q_L=-30~\text{MeV}/c$. 
The $Q_L$ distribution of potential $\pi^+K^-$ pairs shows a similar behaviour 
\cite{CERN-EP-2017-137}. 
For the final analysis, the DIRAC procedure selects events fulfilling 
the following criteria: 
\begin{equation}\label{eq:7c_critq}
Q_T < 4 \;\text{MeV}/c,\quad |Q_L| < 20 \;\text{MeV}/c \, .
\end{equation}

\section{Data simulation}
\label{sec:simul}

\subsection{Multiple scattering simulation}
\label{ssec:MSc}

\begin{figure}[h]
\begin{center}
\includegraphics[width=\linewidth]{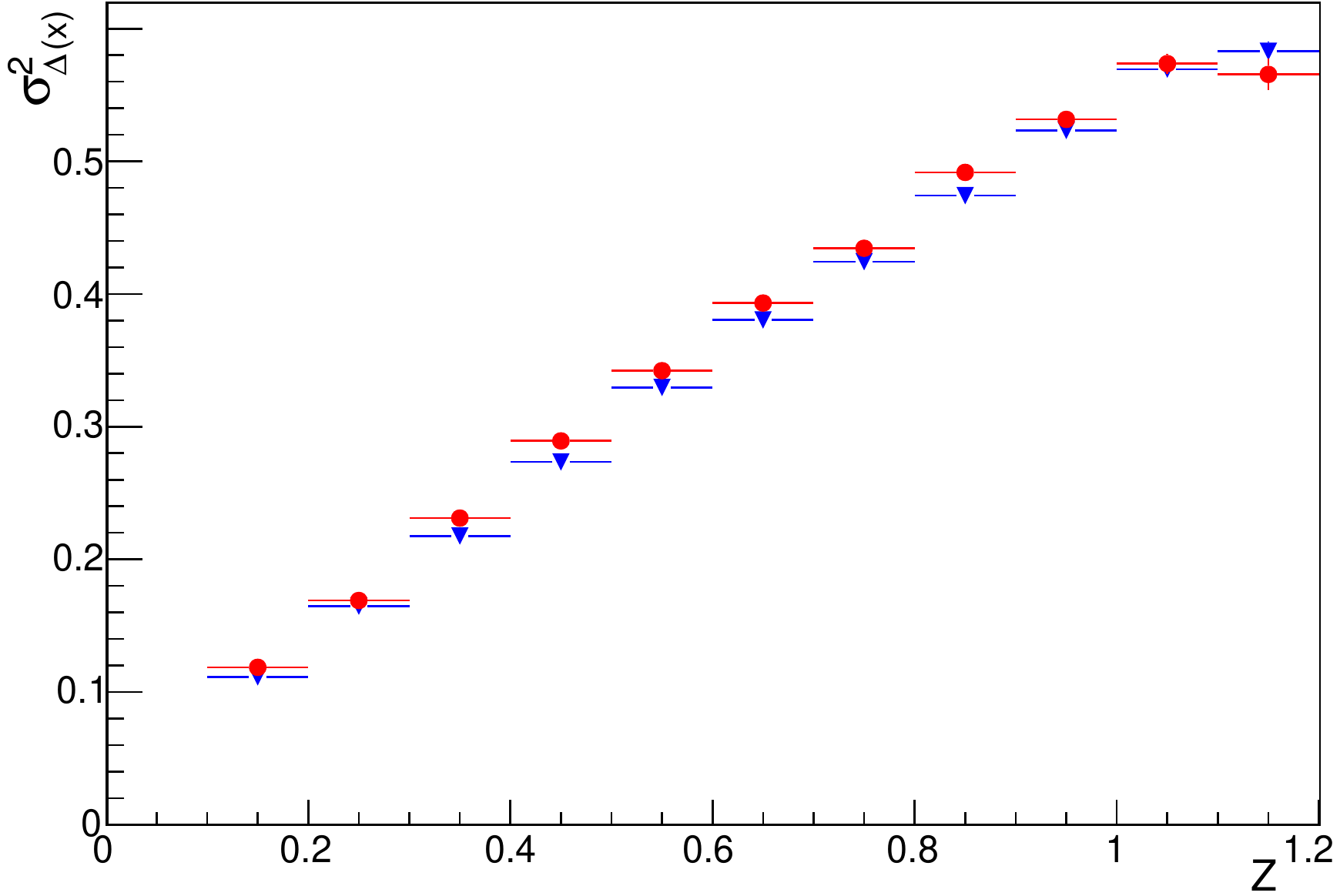}
\end{center}
\caption{$X$ vertex resolution $\sigma^2_{\Delta x}$ in $\rm{cm}^2$ as a
function of
$Z = 1/(p_1\cdot \beta_1)^2 + 1/(p_2\cdot \beta_2)^2$. Experimental
data --- blue triangle, MC data --- red bullet.}
\label{fig:6_4}
\end{figure}

The DIRAC setup as a magnetic vacuum spectrometer has been designed to avoid 
as much as possible distortions of particle momenta by multiple scattering. 
Since particles are scattered in the detector planes, it is essential to
simulate
and reproduce the effect of multiple scattering with a precision better than 
1\%. A detailed study of multiple scattering has already been performed in 
the past \cite{GORC07,msn} and been updated \cite{msn_new} including 
a new evaluation of thickness and density of the SFD material and additionally 
cutting on $|Q_X|$ and $|Q_Y| < 4~\text{MeV}/c$. This cut has been performed 
by the trigger for RUN2 and RUN3 allowing a more accurate comparison between 
data and MC simulation in this region. Prompt $\pi \pi$ pairs were used 
in order to check the correctness of the multiple scattering description in 
the simulation. The events were reconstructed, and tracks of 
positively and negatively charged particles are extrapolated to the target
plane: $x_2$ ($x_1$) and $y_2$ ($y_1$) are the $\pi^{+}$ ($\pi^{-}$) track 
coordinates on the target plane. The experimental error in the track 
measurement and multiple scattering determine the width of 
$\Delta x =x_2 - x_1$ and 
$\Delta y =y_2 - y_1$, called vertex resolution. The vertex resolution as 
a function of the total momentum was studied for particle track pairs with 
momenta $p_1$, $p_2$ and velocities $\beta_1$, $\beta_2$ by using 
the following parameterisation ($X$ direction): 
$$\sigma^2_{\Delta x} = c_1^2 + \frac{s_1^2}{(p_1\cdot \beta_1)^2} + 
                        c_2^2 + \frac{s_2^2}{(p_2\cdot \beta_2)^2}. $$
Here, $c_1$ and $c_2$ account for the momentum independent contribution to
 $\sigma$ (width) of the $x_1$ and $x_2$ distributions 
 and terms with $s_1$ and $s_2$ 
account for the momentum dependent contributions to $\sigma$. 
Assuming $c_1=c_2=c$ and 
$s_1=s_2=s$, one gets 
$$ \sigma^2_{\Delta x} = 2 \cdot c^2 + \left( \frac{1}{(p_1\cdot \beta_1)^2} +
\frac{1}{(p_2\cdot \beta_2)^2}\right)  \cdot s^2
= 2 \cdot c^2 + Z \cdot s^2.$$
Fig.~\ref{fig:6_4} shows for RUN2 a perfect agreement between data and MC 
for the $X$ coordinate, the same is valid for the $Y$ coordinate. 
This procedure, performed for every year of data taking, yields a good 
agreement with the simulation.

\subsection{SFD response}
\label{ssec:SFD}

Track pairs contributing to the signal are characterised by
different opening angles, including very small ones. Therefore, 
it is essential that the SFD detector, which reconstructs 
upstream tracks, is well described in the simulation.
\begin{figure}[h]
\begin{center}
\includegraphics[width=.49\linewidth]{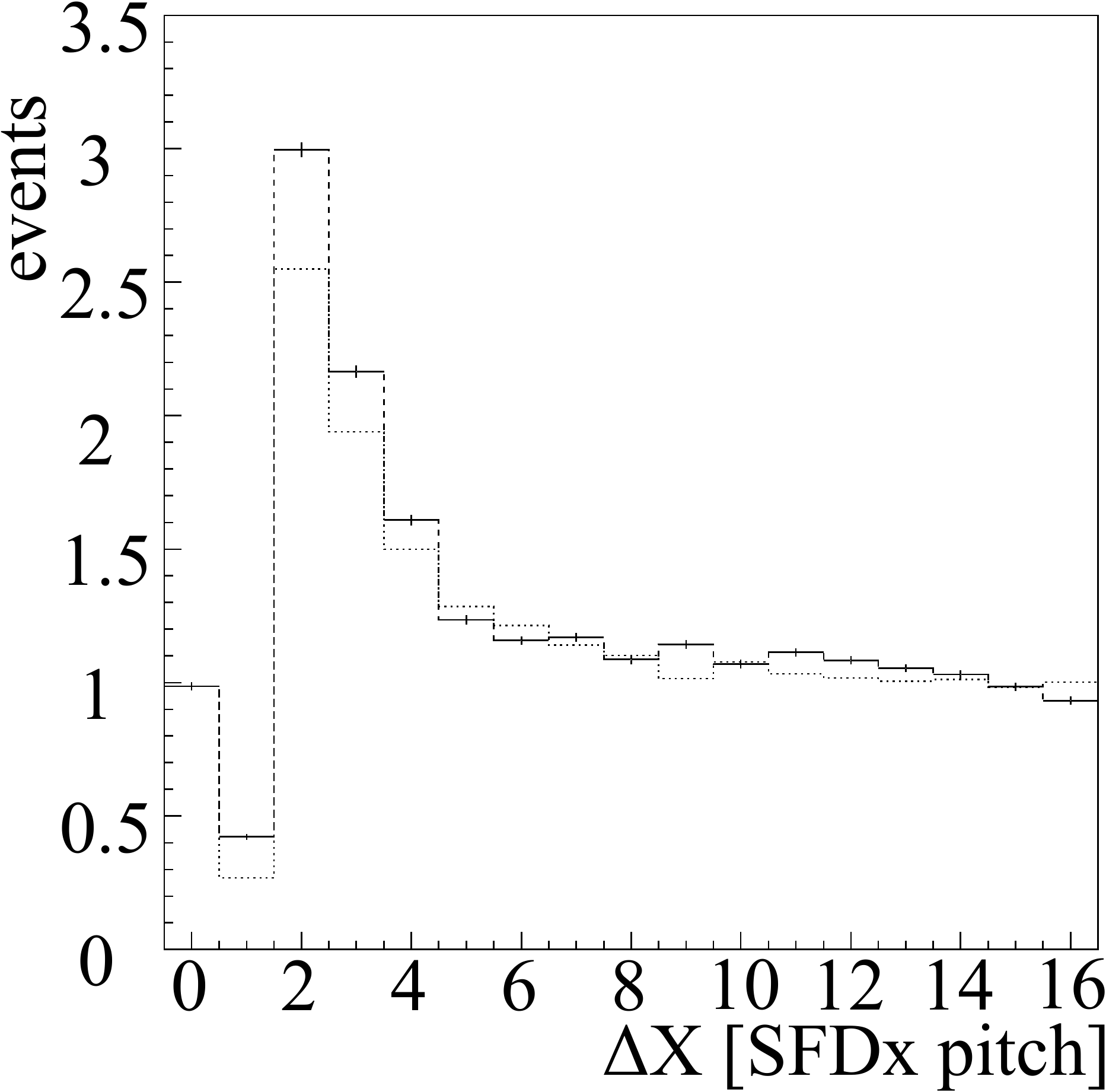}\hfil
\includegraphics[width=.49\linewidth]{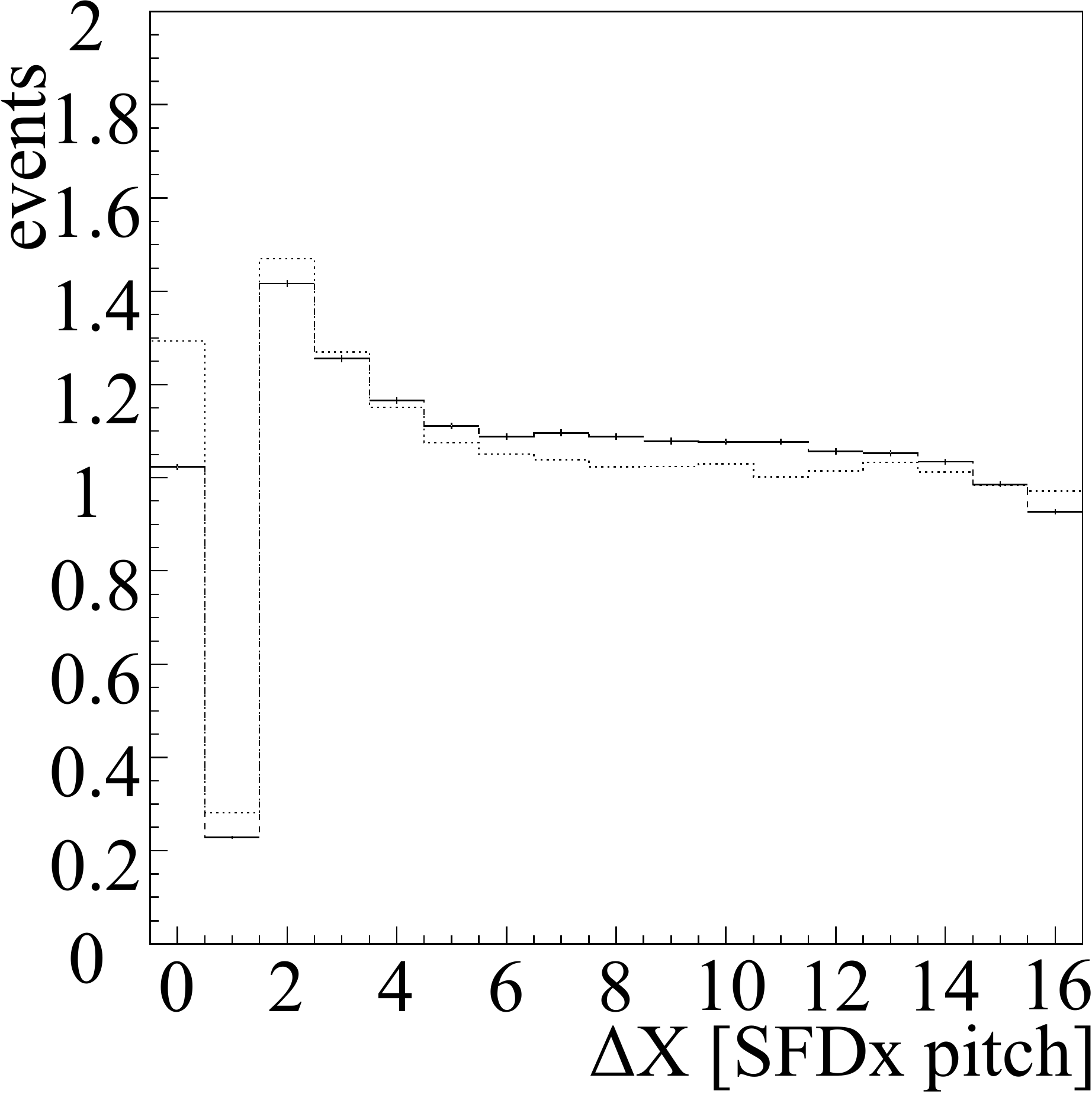}
\end{center}
\caption{Left: $\Delta n$ distribution in SFDx for track pairs with 
small $\Delta n$ in $Y$ ($\Delta n_Y < 3$). 
Right: $\Delta n$ distribution in SFDx without any constraint in $Y$. 
Solid line: experimental data; dotted line: MC data.}
\label{fig:6_5}
\end{figure}

From the $\pi^+\pi^-$ sample outside the signal region 
$(|Q_L| > 10~\text{MeV}/c)$, track pairs with small opening angles 
(small distance between SFD hits) were chosen  for comparison  
of experimental and simulated data. To compare experimental and MC data, 
the events were classified depending on the distance $\Delta n$ 
between the tracks in SFD column number. As an example, 
Fig.~\ref{fig:6_5} (left) shows the $\Delta n_X$ distribution 
of very close tracks in $Y$ ($\Delta n_Y < 3$) and 
Fig.~\ref{fig:6_5} (right) the $\Delta n_X$ distribution without 
any constraint in $Y$ for data of RUN3. (For more details and data 
from the other runs, see \cite{note1603}.) The remaining difference 
between experimental and MC data (Fig.~\ref{fig:6_5}) is corrected 
with weights, which depend on the combination of $\Delta n$ in 
all 3 planes, providing equal $\Delta n$ distributions. 

The new MC simulation takes into account: 
hit efficiency, electronic and photomultiplier noise, 
cluster size associated with a track and background hits 
from beam pipe tracks or from particle scattering in 
the shielding around the detector. These parameters have been 
evaluated for every run, and the comparison between data and 
simulation is satisfactory. The SFD multiplicities in 
the 3 planes are shown in Table~\ref{tab:2} for 
experimental and in Table~\ref{tab:3} for MC data. 

\begin{table}[htbp]
\caption{SFD hit multiplicity for experimental data.}
\label{tab:2}
\begin{ruledtabular}
\begin{tabular}{cccc}
 RUN &  SFDx  &   SFDy  &  SFDu \\ 
\hline
  1  & -- &  $3.4 \pm 0.7$ &  $3.0 \pm 0.7$  \\
\hline   
  2 & $3.6 \pm 0.8$ & $4.1 \pm 1.0 $ & $3.6 \pm 0.8 $ \\
\hline   
  3 & $3.3 \pm 0.8$ & $3.7 \pm 0.9$ & $3.2 \pm 0.8  $\\
\hline   
  4 & $2.9 \pm 0.8$ & $3.3 \pm 1.0$ & $3.0 \pm 0.8 $ \\
\end{tabular}
\end{ruledtabular}
\end{table}

\begin{table}[htbp]
\caption{SFD hit multiplicity for MC data.}
\label{tab:3}
\begin{ruledtabular}
\begin{tabular}{cccc}
 RUN &  SFDx  &   SFDy & SFDu \\ 
 \hline
 1 & -- &  $3.5 \pm 0.6$ & $ 3.4 \pm 0.6$  \\
\hline   
 2 & $3.8 \pm 0.6$ & $4.0 \pm 0.6$ & $3.7 \pm 0.6 $ \\
\hline   
 3 & $3.3 \pm 0.6$ & $3.6 \pm 0.6$ & $3.3 \pm 0.6 $ \\
\hline   
 4 & $3.1 \pm 0.8$ & $3.4 \pm 1.0$ & $3.0 \pm 0.8 $  \\
\end{tabular}
\end{ruledtabular}
\end{table}

\subsection{Momentum resolution} 
\label{ssec:momentum}

\begin{figure*}[htb]
\includegraphics[width=0.47\linewidth]{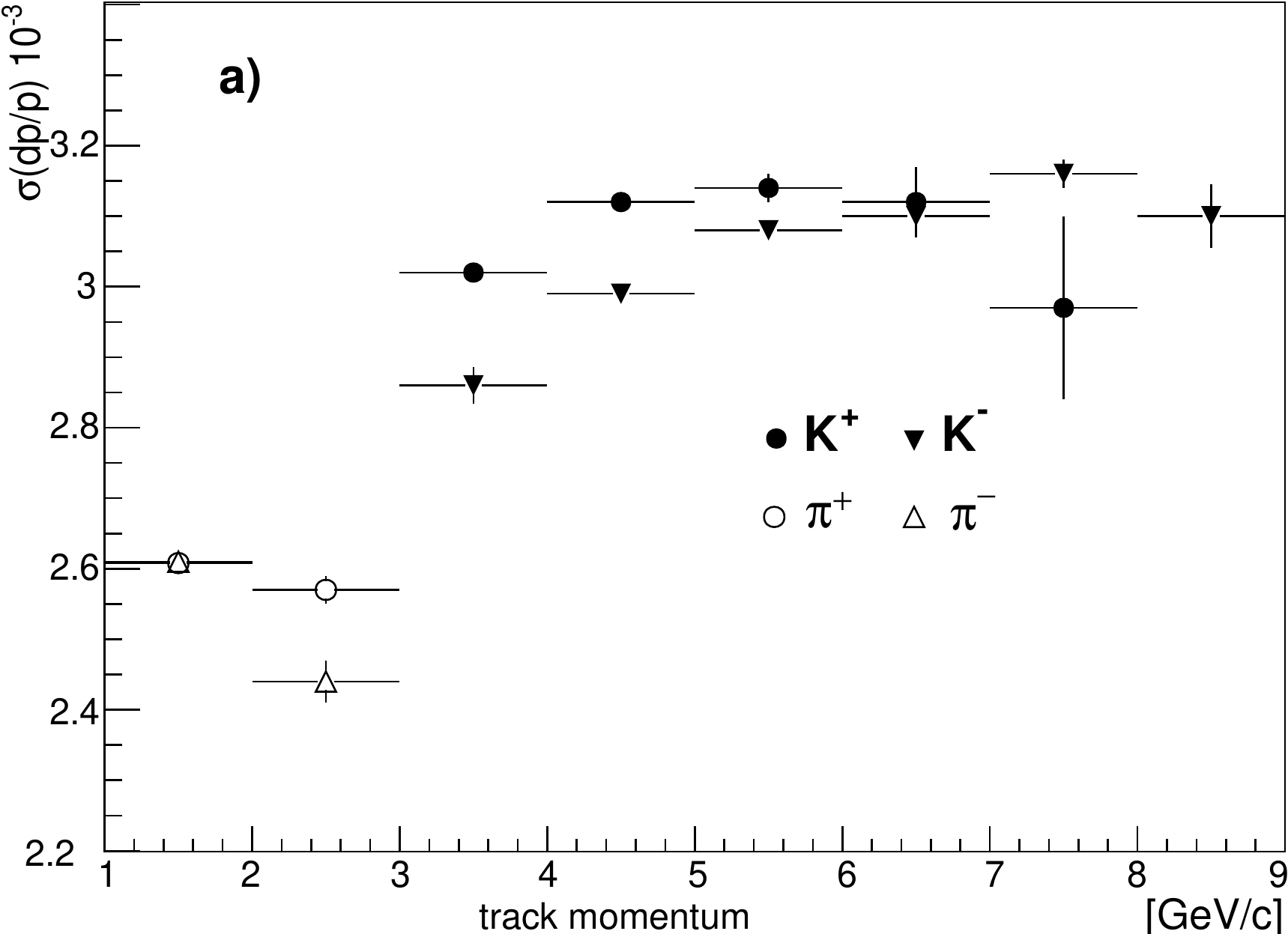}\hfil
\includegraphics[width=0.47\linewidth]{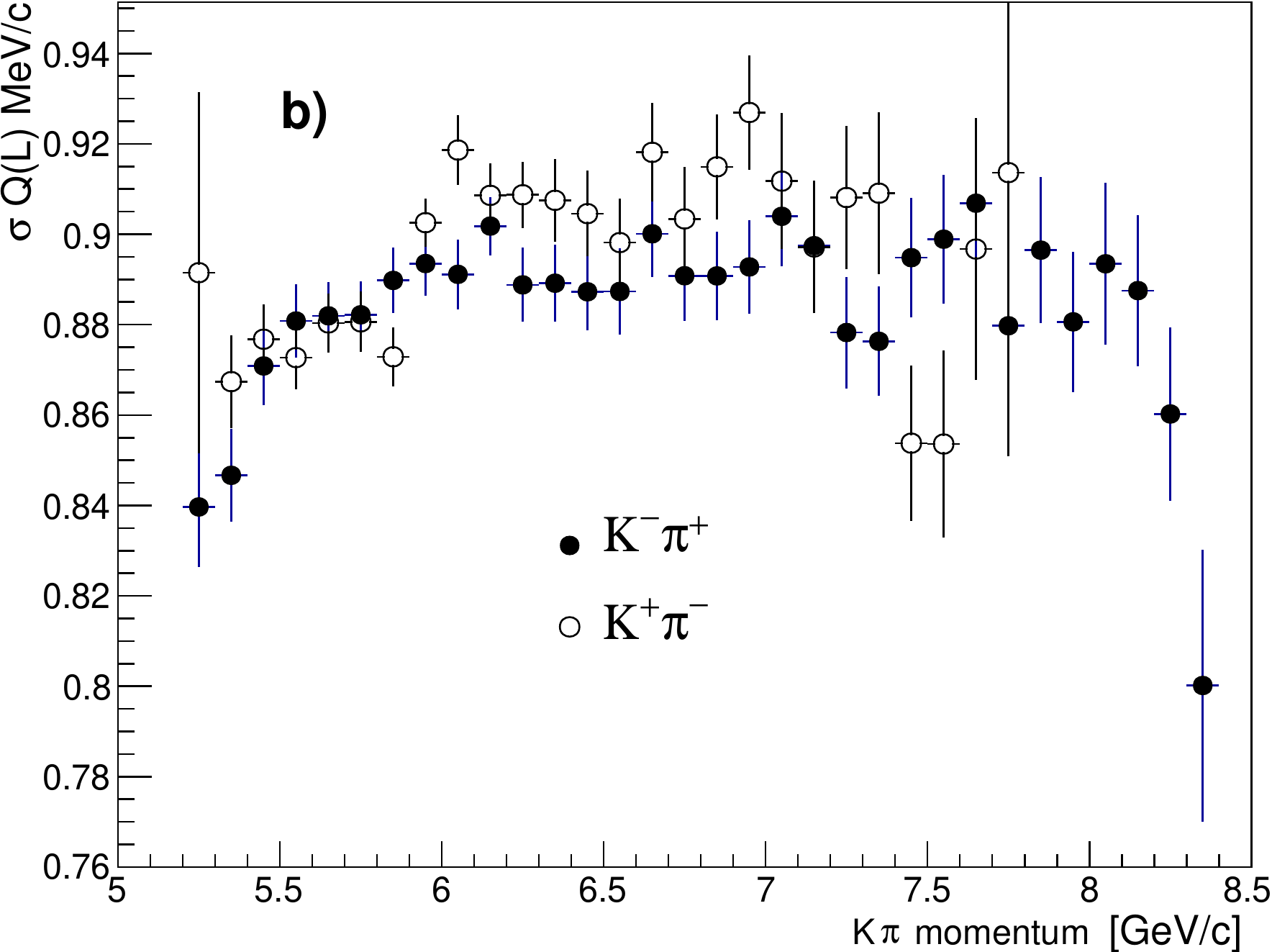}\\
\includegraphics[width=0.47\linewidth]{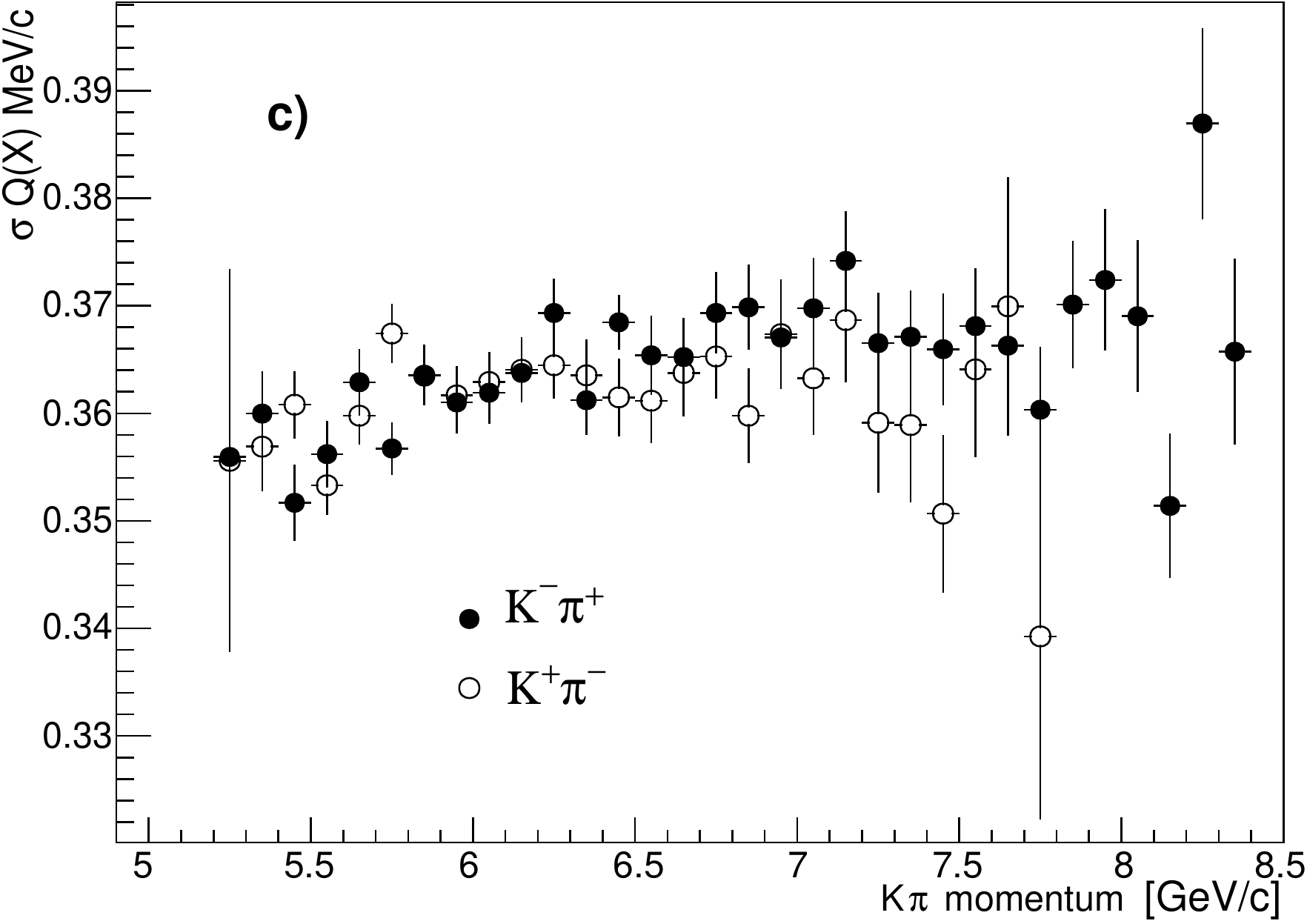}\hfil
\includegraphics[width=0.47\linewidth]{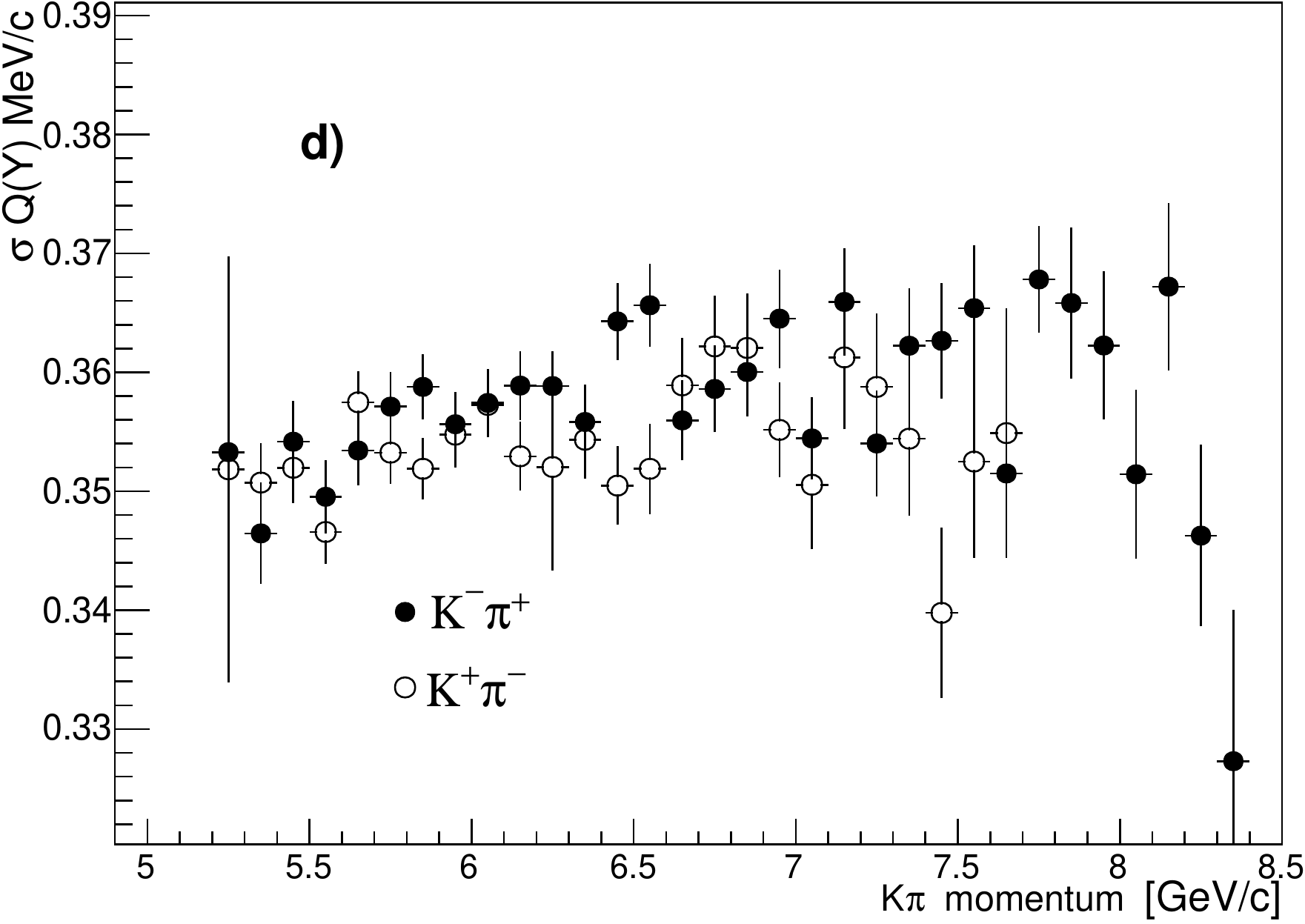}
\caption{Momentum resolution (a) as a function of the lab particle momentum 
and resolution of the relative momentum components 
$Q_{L}$ (b), $Q_{X}$ (c) and $Q_{Y}$ (d) as a function of 
the total lab momentum of $K^{-}\pi^{+}$ (black) 
and $K^{+}\pi^{-}$ (white).}
\label{fig:6_7}
\end{figure*}

Using simulated $\pi K$ events, the momentum resolution is evaluated by means 
of the expression $\delta_{p} = (p_\mathrm{gen} - p_\mathrm{rec}) /
p_\mathrm{gen}$,
where $p_\mathrm{gen}$ and $p_\mathrm{rec}$ are the generated and reconstructed 
momenta, respectively. The additional momentum smearing was taken 
into account (Section~\ref{ssec:geom}). The resulting $\delta_{p}$ 
distributions were fitted with a Gaussian, and the standard deviations  
$\sigma$ of the distributions as a function of the particle momentum 
$p_{rec}$ are presented in Fig.~\ref{fig:6_7}a. In the range from 
1 to 8~$\text{GeV}/c$, the DIRAC spectrometer reconstructs lab momenta with 
a relative precision between $2.4 \cdot 10^{-3}$ and $3.2 \cdot 10^{-3}$. 
The resolution of the relative momentum components $Q_{L}$, $Q_{X}$ 
and $Q_{Y}$ are obtained by MC simulation in the same 
approach as for the momentum resolution. The results for RUN4 are shown 
in Fig.~\ref{fig:6_7}. For the other runs, the resolutions are similar. 
	
\subsection{Simulation of atomic, Coulomb and non-Coulomb $\pi K$ pair
production} 
\label{ssec:pairs}

Non-Coulomb $\pi K$ pairs, not affected by FSI, show uniform distributions in 
the c.m. relative momentum projections, whereas Coulomb pairs, exposed to 
Coulomb FSI, show distributions corresponding to uniform distributions 
modified by the Gamov-Sommerfeld-Sakharov factor (\ref{eq:cross_sect_C}). 
The MC distributions of the lab pair momentum are based on 
the experimental momentum distributions \cite{GORC10}. 
The $\pi^+K^-$ were simulated according to $dN/dp=e^{-0.50p}$ and 
the $\pi^-K^+$ pairs according to $dN/dp=e^{-0.89p}$, where 
$p$ is the lab pair momentum in $\text{GeV}/c$. After comparing  
the experimental with the MC distribution analyzed by 
the DIRAC program ARIANE, the simulated distributions were modified 
by applying a weight function in order to fit the experimental data. 
The lab momentum spectrum of simulated atoms is the same as for 
Coulomb pairs (\ref{eq:prod}). Numerically solving the transport 
equations (Section~\ref{sec:interact}), allows to obtain the distributions of 
the atom breakup points in the target and of the atomic states at the breakup. 
The latter distribution defines the original c.m. relative momenta $q$ of 
the produced atomic pairs. The initial spectra of MC atomic, Coulomb and 
non-Coulomb pairs have been generated by the DIPGEN code \cite{DIPGEN}. 
Then, these pairs propagate through the setup according to the detector 
simulation program GEANT-DIRAC and get analyzed by ARIANE.

The description of the charged particle propagation takes into account 
(a) multiple scattering in the target, detector planes and setup partitions, 
(b) the response of all detectors, 
(c) the additional momentum smearing (Section~\ref{ssec:geom}) and 
(d) the results of the SFD response analysis (Section~\ref{ssec:SFD}) 
influencing the $Q_T$ resolution.

The propagation of $A_{\pi K}$ through the target is simulated by 
the MC method. The total amount of atomic pairs is $n_A^{MC}(0)$. 
The full number of simulated Coulomb pairs in the same setup acceptance 
is $N^{MC}_C(0)$, and the amount of Coulomb pairs with relative momenta 
$q < 3.12~\text{MeV}/c$ (\ref{eq:number_A}) is $N^{MC}_C(K)$. 
These numbers are used for calculating the atom breakup probabilities.

\section{Data analysis}
\label{sec:analysis}

\subsection{Number of $\pi^-K^+$ and $\pi^+K^-$ atoms and atomic pairs} 
\label{ssec:nA}

The analysis of $\pi K$ data is similar to the $\pi^+\pi^-$ analysis as 
presented in \cite{ADEV11}. For events with $Q_T < 4~\text{MeV}/c$ 
and $|Q_L| < 20~\text{MeV}/c$ (\ref{eq:7c_critq}), 
the experimental distributions of $Q$ ($N(Q_i)$) and 
of its projections have been fitted for each run and 
each $\pi K$ charge combination by simulated distributions of 
atomic ($n^{MC}_A(Q_i)$), Coulomb ($N^{MC}_{C}(Q_i)$) and 
non-Coulomb ($N^{MC}_{nC}(Q_i)$) pairs. The admixture of accidental pairs 
has been subtracted from the experimental distributions, using the difference 
of the particle production times (Section~\ref{ssec:Ev_sel}). The distributions 
of simulated events are normalized to 1 by integrating them 
($n^{MC}_A$, $N^{MC}_C$ and $N^{MC}_{nC}$).
In the experimental distributions, the numbers of 
atomic ($n_A$), Coulomb ($N_C$) and non-Coulomb ($N_{nC}$) pairs 
are free fit parameters in the minimizing expression:
\begin{widetext}
\begin{equation}\label{eq:fitdata-chi2}
\chi^2 =  \sum_i \frac{\left( N(Q_i) -
n_A \cdot n^{MC}_A(Q_i) -
N_C \cdot N^{MC}_{C}(Q_i) -
N_{nC} \cdot N^{MC}_{nC}(Q_i) \right)^2}
{\sigma^{2}_{N(Q_i)}} \,.
\end{equation}
\end{widetext}
The sum of these parameters is equal to the number of analyzed events. 
The fitting procedure takes into account the statistical errors of 
the experimental distributions. The statistical errors of the MC distributions 
are more than one order less than the experimental ones. 

\begin{figure}[htbp]
\begin{center}
\includegraphics[width=\linewidth]{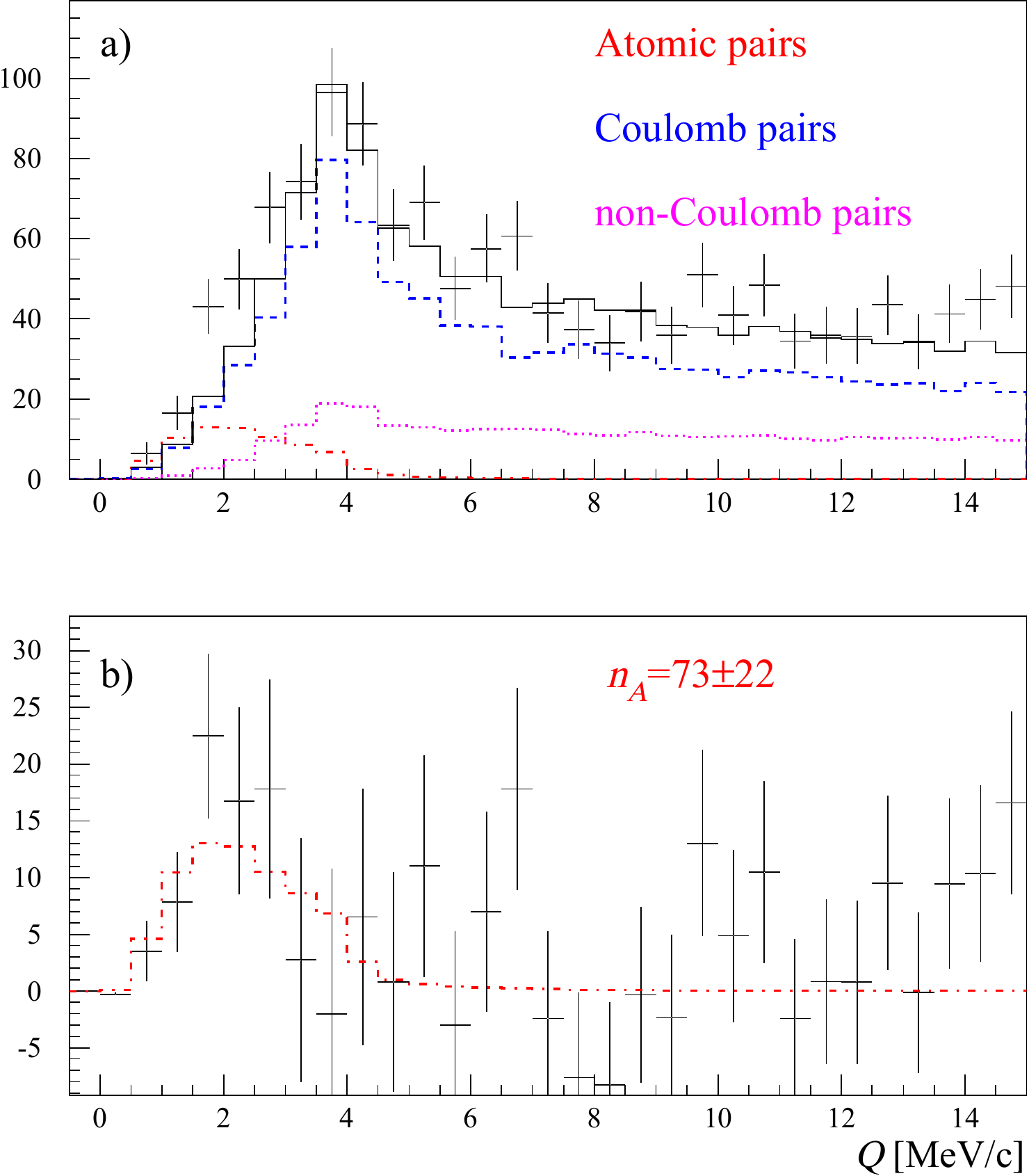}
\caption{a) Experimental distribution of $\pi^-K^+$ and $\pi^+K^-$ pairs 
(points with error bars) for the platinum (Pt) target fitted by a sum of 
simulated distributions of ``atomic'', ``Coulomb'' and ``non-Coulomb'' pairs. 
The background distribution of free (``Coulomb'' and ``non-Coulomb'') pairs 
is shown as black line; 
b) Difference distribution between the experimental and simulated 
free pair distributions compared with the simulated 
distribution of ``atomic pairs''.} 
\label{fig:Q_Pt}
\end{center}
\end{figure}

\begin{figure}[htbp]
\begin{center}
\includegraphics[width=\linewidth]{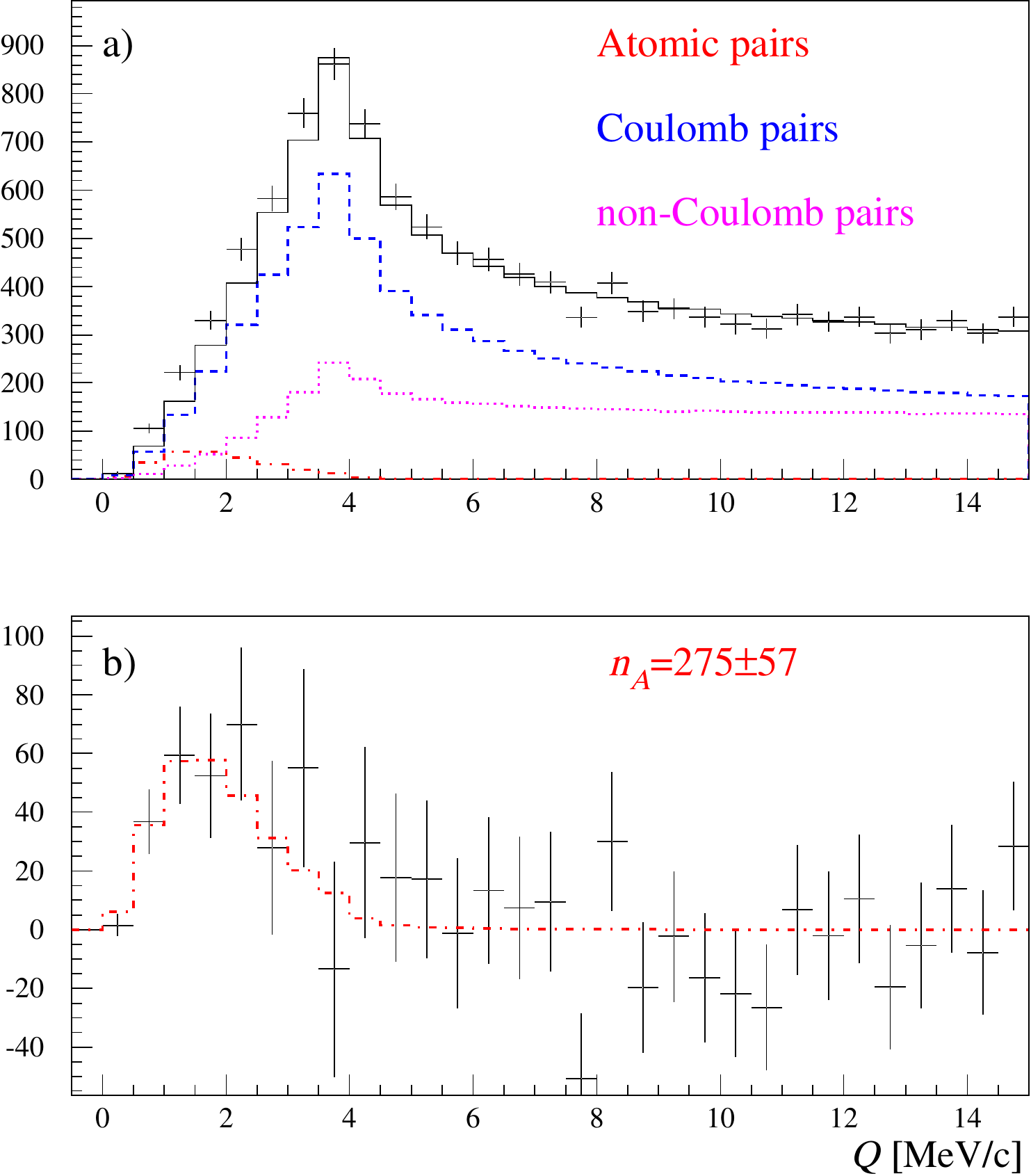}
\caption{Experimental distribution of $\pi^-K^+$ and $\pi^+K^-$ pairs 
for nickel (Ni) target analogous to Fig.~\ref{fig:Q_Pt}.}
\label{fig:Q_Ni}
\end{center}
\end{figure}

Fig.~\ref{fig:Q_Pt}a presents the experimental and simulated $Q$ distributions
of $\pi K$ pairs for the data obtained from the Pt target and 
Fig.~\ref{fig:Q_Ni}a for Ni data. One observes an excess of events above the 
sum of Coulomb and 
non-Coulomb pairs in the low $Q$ region, where atomic pairs are expected: 
these excess spectra are shown in Figs.~\ref{fig:Q_Pt}b and \ref{fig:Q_Ni}b 
together with the simulated distribution of atomic pairs. 
The numbers of atomic pairs, found in the Pt and Ni target data, are 
$n_A(\text{Pt})=73 \pm 22$ ($\chi^2/n = 40/36$, $n =$ number of degrees of
freedom)
and $n_A(\text{Ni})=275 \pm 57$ ($\chi^2/n = 40/37$). Comparing the 
experimental and simulated distributions demonstrates good agreement.
 
The same analysis was performed for $\pi^-K^+$ and $\pi^+K^-$ pairs, 
separately. For the Pt target, the numbers of $\pi^-K^+$ and $\pi^+K^-$ 
atomic pairs are $n^{\pi^-K^+}_A(\text{Pt})=57 \pm 19$ 
($\chi^2/n = 40/36$) and $n^{\pi^+K^-}_A(\text{Pt})=16 \pm 12$ 
($\chi^2/n = 41/36$), and for Ni, the corresponding numbers are 
$n^{\pi^-K^+}_A(\text{Ni})=186 \pm 48$ ($\chi^2/n = 33/37$) and 
$n^{\pi^+K^-}_A(\text{Ni})=90 \pm 30$ ($\chi^2/n = 39/37$). 
The experimental ratios between the two types of atom production are 
$3.5 \pm 2.7$ for Pt and $2.07 \pm 0.87$ for Ni. Corrected by 
the difference of their detection efficiencies, these ratios result in 
$R^{\pi^-K^+}_{\pi^+K^-}(\text{Pt})=3.2 \pm 2.5$ and 
$R^{\pi^-K^+}_{\pi^+K^-}(\text{Ni})= 2.5 \pm 1.0$, compatible with 
$2.4$ as calculated in the framework of FRITIOF \cite{GORC16}. 
Tables \ref{tab:nA_Pt} and \ref{tab:nA_Ni} present these data, comparing them 
with the results of the $|Q_L|$ and the 2-dimensional ($|Q_L|$,$Q_T$) analyzes.
The results of the $Q$ and ($|Q_L|$,$Q_T$) analyzes are in good agreement, and 
the 1-dimensional $|Q_L|$ analysis does not contradict the values obtained in 
the other two statistically more precise analyzes. 
 
\begin{table}[htb]
\caption{$\pi^-K^+$ and $\pi^+K^-$ data for the Pt target: 
atomic pair numbers $n_A$ and ratio $R^{\pi^-K^+}_{\pi^+K^-}$ 
as obtained by analyzing the 1-dimensional $Q$ and $|Q_L|$ distributions 
and the 2-dimensional ($|Q_L|$,$Q_T$) distribution. 
Only statistical errors are given.}
\label{tab:nA_Pt}
\begin{ruledtabular}
\begin{tabular}{ccccc}
Analysis & $n_A$ & $n^{\pi^-K^+}_A$ & $n^{\pi^+K^-}_A$ &
$R^{\pi^-K^+}_{\pi^+K^-}$ \\
&($\chi^2/n$) & ($\chi^2/n$) & ($\chi^2/n$) &  \\
\hline
$Q$ & $73 \pm 22$ & $57 \pm 19$ & $16 \pm 12$ & $3.2 \pm 2.5$ \\
    & (40/36) &  (40/36) & (41/36) & \\
\hline
$|Q_L|$ & $73 \pm 31$& $61 \pm 27$ & $12 \pm 16$ & $4.7 \pm 6.6$ \\
& (37/37) & (40/37) & (28/37) & \\
\hline
$|Q_L|,Q_T$ & $71 \pm 21$ & $65 \pm 18$ & $6 \pm 11$& $10 \pm 20$ \\
& (169/154) & (159/151) & (102/135) & \\
\end{tabular}
\end{ruledtabular}
\end{table}
\begin{table}[htb]
\caption{$\pi^-K^+$ and $\pi^+K^-$ data for the Ni targets: 
atomic pair numbers $n_A$ and ratio $R^{\pi^-K^+}_{\pi^+K^-}$ 
analogous to Table \ref{tab:nA_Pt}.}
\label{tab:nA_Ni}
\begin{ruledtabular}
\begin{tabular}{ccccc}
Analysis & $n_A$ & $n^{\pi^-K^+}_A$ & $n^{\pi^+K^-}_A$ &
$R^{\pi^-K^+}_{\pi^+K^-}$ \\
& ($\chi^2/n$) & ($\chi^2/n$) & ($\chi^2/n$) & \\
\hline
$Q$ & $275 \pm 57$ & $186 \pm 48$ & $90 \pm 30$ & $2.5 \pm 1.0$ \\
& (40/37) & (33/37) & (39/37) & \\
\hline
$|Q_L|$ & $157 \pm 87$ & $103 \pm 74$ & $55 \pm 45$ & $2.3 \pm 2.5$ \\
& (56/37) & (52/37) & (32/37) & \\
\hline
$|Q_L|,Q_T$ & $243 \pm 56$ & $171 \pm 47$ & $72 \pm 30$ & $2.8 \pm 1.4$ \\
& (225/157) & (226/157) & (157/157) & \\
\end{tabular}
\end{ruledtabular}
\end{table}

The efficiency of atomic pair recording is evaluated from 
the simulated data as ratio of the MC atomic pair number $n_A^{MC}$, 
passed the corresponding cuts - in each of the above analysis - to 
the full number of generated atomic pairs: 
$\varepsilon_A=n_A^{MC}/n_A^{MC}(0)$ (Section \ref{ssec:pairs}). 
The full number of atomic pairs, that corresponds to 
the experimental value $n_A$, is given by $n_A/\varepsilon_A$. 
In the same way, the efficiency of Coulomb pair recording is 
$\varepsilon_C=N_C^{MC}/N_C^{MC}(0)$ and the full number of 
Coulomb pairs $N_C/\varepsilon_C$. This number allows to calculate 
the number $N_A$ of atoms produced in the target, using  
the theoretical ratio $K$ (\ref{eq:number_A}) and the simulated efficiency 
$\varepsilon_K=N_C^{MC}(K)/N_C^{MC}(0)$ of the cut $q < 3.12~\mathrm{MeV}/c$ 
for Coulomb pairs: $N_A=K\cdot\varepsilon_K\cdot N_C/\varepsilon_C$. 
Thus, the atom breakup probability $P_\mathrm{br}$ is expressed via 
the fit results $n_A$, $N_C$ and the simulated efficiencies as:
\begin{equation}\label{eq:pbr}
P_\mathrm{br} = \frac{\frac{n_A}{\varepsilon_A}}
{K \cdot \varepsilon_K \frac{N_C}{\varepsilon_C}} \, .
\end{equation}

Table~\ref{tab:Pbr} contains the $P_\mathrm{br}$ values obtained in the $Q$ 
and ($|Q_L|$,$Q_T$) analyzes. 
\begin{table}[htb]
\caption{Experimental $P_\mathrm{br}$ from $Q$ and $(|Q_L|,Q_T)$ analyzes. 
Only statistical uncertainties are cited.}
\label{tab:Pbr}
\begin{ruledtabular}
\begin{tabular}{ccccc}
 Data & RUN & Target (\textmu{}m) & $P^{Q}_{br}$ & $P^{|Q_L|,Q_T}_{br}$ \\
\hline
 $\pi^+K^-$ & 1 & Pt (25.7) &  $1.2 \pm 1.3$  & $0.27 \pm 0.56$ \\
 $\pi^+K^-$ & 2 & Ni  (98)  & $0.53 \pm 0.39$ & $0.42 \pm 0.38$ \\
 $\pi^+K^-$ & 3 & Ni (108)  & $0.29 \pm 0.20$ & $0.33 \pm 0.24$ \\
 $\pi^+K^-$ & 4 & Ni (108)  & $0.33 \pm 0.22$ & $0.21 \pm 0.20$ \\
\hline
 $\pi^-K^+$ & 1 & Pt (25.7) & $1.09 \pm 0.52$ & $1.44 \pm 0.59$ \\
 $\pi^-K^+$ & 2 & Ni  (98)  & $0.32 \pm 0.20$ & $0.44 \pm 0.22$ \\
 $\pi^-K^+$ & 3 & Ni (108)  & $0.23 \pm 0.16$ & $0.16 \pm 0.15$ \\
 $\pi^-K^+$ & 4 & Ni (108)  & $0.41 \pm 0.17$ & $0.34 \pm 0.16$ \\
\hline
$\pi^+K^-\&K^+\pi^-$ & 1 & Pt, 25.7 & $1.11 \pm 0.48$ & $0.83 \pm 0.41$ \\ 
\end{tabular}
\end{ruledtabular}
\end{table}

\subsection{Systematic errors} 
\label{ssec:systematic}

Different sources of systematic errors were investigated. Most of them 
arise from differences in the shapes of experimental and MC 
distributions for atomic, Coulomb and, to a much lesser extent, 
for non-Coulomb pairs. The shape differences induce a bias in 
the values of the fit parameters $n_A$ and $N_C$, leading to 
systematic errors of the atomic pair number and finally of 
the probability $P_\mathrm{br}$. In the following, a list of 
the different sources is presented:

\begin{itemize}
\item Resolution over particle momentum of the simulated events is modified 
by the $\Lambda$ width correction (Section \ref{ssec:geom}). The parameter $C$, 
used for additional smearing of measured momenta, is defined with finite
accuracy, resulting in a possible difference in resolution of experimental 
and simulated data over $Q_L$.

\item Multiple scattering in the targets (Pt and Ni) provides 
a major part of the $Q_T$ smearing. The average multiple scattering angle 
is known with 1\% accuracy. This uncertainty induces a systematic error 
due to different resolutions over $Q_T$ for experimental and simulated data. 

\item SFD simulation procedure as described in Section \ref{ssec:SFD} 
corrects a residual difference with weights, depending on the distances 
between particles in the three SFD planes. These weights are estimated
by a separate procedure resulting in a systematic error. 

\item Coulomb pair production cross section increases at low $q$ according to 
$A_C(q)$ (\ref{eq:cross_sect_C}) assuming a pointlike pair production region. 
Typical sizes of production regions from medium-lived particle decays 
[($30 \div 40$)~fm] are smaller than the Bohr radius (such pairs undergo 
Coulomb FSI), but not pointlike. In order to check finite size effects 
due to the presence of medium-lived particles ($\omega$, $\phi$), 
non-pointlike particle pair sources are investigated, and correlation functions 
for the different pair sources calculated \cite{LEDN08}. 
The final correlation function, considering the sizes of 
the pair production regions, has some uncertainty due to 
limited accurate fractions of the different $\pi K$ sources.  

\item Uncertainties in the measurement of $\pi^-K^+$ and $\pi^+K^-$ 
pair lab momentum spectra and the relation between these uncertainties 
and the systematic errors of the atomic pair measurement are described 
in \cite{note1306}.
There is a mechanism that increases the influence of the bias between 
experimental and simulated distributions for $\pi K$ compared to $\pi\pi$. 
For detected small $Q$ $\pi K$ pairs, kaons have lab momenta 
$\sim 3.5$ times higher than pions, $(4 \div 6)~\text{GeV}/c$ compared to 
$(1.2 \div 2)~\text{GeV}/c$. The spectrometer acceptance as a function of 
lab momentum strongly decreases at momenta higher than 3~$\text{GeV}/c$. 
As a result, kaons with lower momenta are detected more efficiently. 
In the pair c.m. system, this corresponds to $Q_L < 0$ for $\pi^-K^+$ pairs 
as illustrated in Fig.~\ref{fig:kpi-selection}c. 
For $\pi\pi$, the corresponding distributions consist of the flat horizontal 
background of non-Coulomb pairs and symmetric peak of Coulomb and atomic pairs.  
The observed slope for $\pi K$ in $Q_L$ distribution is non-linear, that 
transforms to a non-linear background behavior in $|Q_L|$.
Thus, the quality of separation between Coulomb and non-Coulomb pairs
becomes more sensitive to the accuracy of simulated distributions. 

\item Uncertainty in the lab momentum spectrum of background pairs results in 
a similar effect as the uncertainties of $\pi^-K^+$ and $\pi^+K^-$ spectra. 
Both spectra are measured with a time-of-flight based procedure 
(Section~\ref{sec:backr_subtr}), but as independent parameters. Therefore, 
the uncertainty of the background pairs is assumed to be an independent source 
for systematic errors. 

\item Uncertainty in the $P_\mathrm{br}(\tau)$ relation (Section \ref{ssec:br}).

\end{itemize}

Estimations of systematic errors, induced by different sources, are presented 
in Table~\ref{tab:systPt} for Pt data and Table~\ref{tab:systNi} for Ni data. 
The total errors were calculated as the quadratic sum. The procedure of 
the $\pi K$ atom lifetime estimation described below includes all systematic
errors,
although their contributions are insignificant compared to the statistical
errors.
\begin{table}[htb]
\caption{Estimated systematic errors of $P_\mathrm{br}$ for Pt in $Q$ 
and ($|Q_L|,Q_T$) analyzes.}
\label{tab:systPt}
\begin{ruledtabular}
\begin{tabular}{p{0.65\columnwidth}ll}
\rule{0pt}{2.5ex} Source & $Q$ & $(|Q_L|,Q_T)$ \\
\hline
Uncertainty in $\Lambda$ width correction & 0.011 & 0.073 \\
Uncertainty of multiple scattering in the Pt target & 0.0087 & 0.014 \\
Accuracy of SFD simulation & 0. & 0. \\
Correction of the Coulomb correlation function on finite size production 
region & 0.0001 & 0.0002 \\
Uncertainty in $\pi K$ pair lab. momentum spectrum & 0.089 & 0.25 \\
\raggedright Uncertainty in the laboratory momentum spectrum of background pairs
& 0.22 & 0.21 \\
Uncertainty in the $P_\mathrm{br}(\tau)$ relation & 0.01 & 0.01 \\
\hline
Total & 0.24 & 0.34 \\
\end{tabular}
\end{ruledtabular}
\end{table}

\begin{table}[htb]
\caption{Estimated systematic errors of $P_\mathrm{br}$ for Ni in $Q$ 
and ($|Q_L|,Q_T$) analyzes.}
\label{tab:systNi}
\begin{ruledtabular}
\begin{tabular}{p{0.65\columnwidth}ll}
\rule{0pt}{2.5ex} Source & $Q$ & $(|Q_L|,Q_T)$ \\
\hline
\rule{0pt}{2.5ex}Uncertainty in $\Lambda$ width correction & 0.0006 & 0.0006 \\
Uncertainty of multiple scattering in a Ni target & 0.0051 & 0.0036 \\
Accuracy of SFD simulation & 0.0002 & 0.0003 \\
Correction of the Coulomb correlation function on finite size production 
region & 0.0001 & 0.0000 \\
Uncertainty in $\pi K$ pair lab. momentum spectrum & 0.0052 & 0.0050 \\
\raggedright Uncertainty in the laboratory momentum spectrum of background pairs
& 0.0011 &
0.0011 \\
Uncertainty in the $P_\mathrm{br}(\tau)$ relation & 0.0055 & 0.0055 \\
\hline
Total & 0.0092 & 0.0084 \\
\end{tabular}
\end{ruledtabular}
\end{table}

\subsection{$\pi K$ atom lifetime and $\pi K$ scattering length measurements}
\label{ssec:time}

The $\pi K$ atom breakup probabilities $P_\mathrm{br}= f(\tau,l,Z,p_A)$ 
in the different targets are presented in Section~\ref{ssec:br} and 
have been calculated for the Ni (98~\textmu{}m, 108~\textmu{}m) 
and the Pt (26~\textmu{}m) 
targets. For each target, $P_\mathrm{br}$ is evaluated for $\pi^+K^-$ 
and $\pi^-K^+$ atoms, separately, taking into account their lab momentum 
distributions. For estimating the lifetime of $A_{\pi K}$ in 
the ground state, the maximum likelihood method \cite{daniel08} is 
applied \cite{DN201606}:
\begin{equation}
 L(\tau) = \exp \left(-U^T G^{-1} U/2\right),
\end{equation}
where $U_i=\Pi_i-P_{\mathrm{br},i}(\tau)$ 
is a vector of differences between measured $\Pi_i$ ($P_\mathrm{br}$ in 
Table \ref{tab:Pbr}) and corresponding theoretical breakup probability 
$P_{\mathrm{br},i}(\tau)$ for a data sample~$i$. 
The error matrix of $U$, named $G$, includes statistical ($\sigma_i$) as well 
as systematic uncertainties. Only the term corresponding to the uncertainty in 
the $P_{\mathrm{br}}(\tau)$ relation is considered as correlated between the Ni
and Pt data, which is a conservative approach and overestimates this error. 
The other systematic uncertainties do not exhibit a correlation between 
the data samples from the Ni and Pt targets. On the other hand, 
systematic uncertainties of the Ni data samples are correlated.  

\begin{figure}
\includegraphics[width=0.8\linewidth]{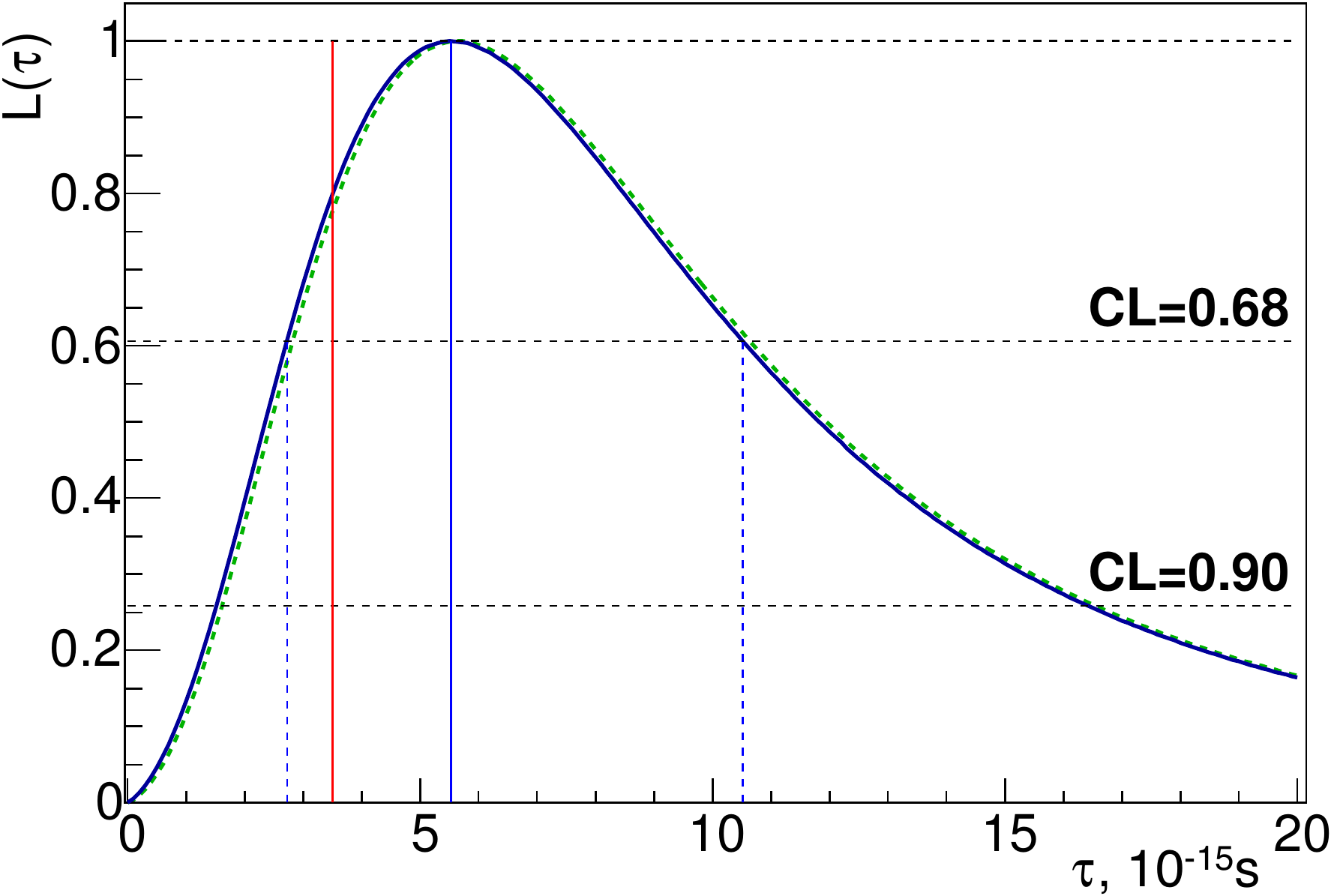}
\caption{Likelihood functions $L(\tau)$ for $Q$ analyzes with
$Q_T<4\:\text{MeV}/c$.
The likelihood functions on the basis of both statistical and systematic errors
(dashed green line) and
  on the basis of only statistical error (solid blue line) are presented. 
  The vertical blue lines indicate the best estimate for $\tau_{\text{tot}}$
  and the corresponding confidence interval. The vertical red line is 
  the theoretical prediction~(\ref{eq:tau35}).
}
\label{fig:likelihoods}
\end{figure}

\begin{figure}
\includegraphics[width=0.9\linewidth]{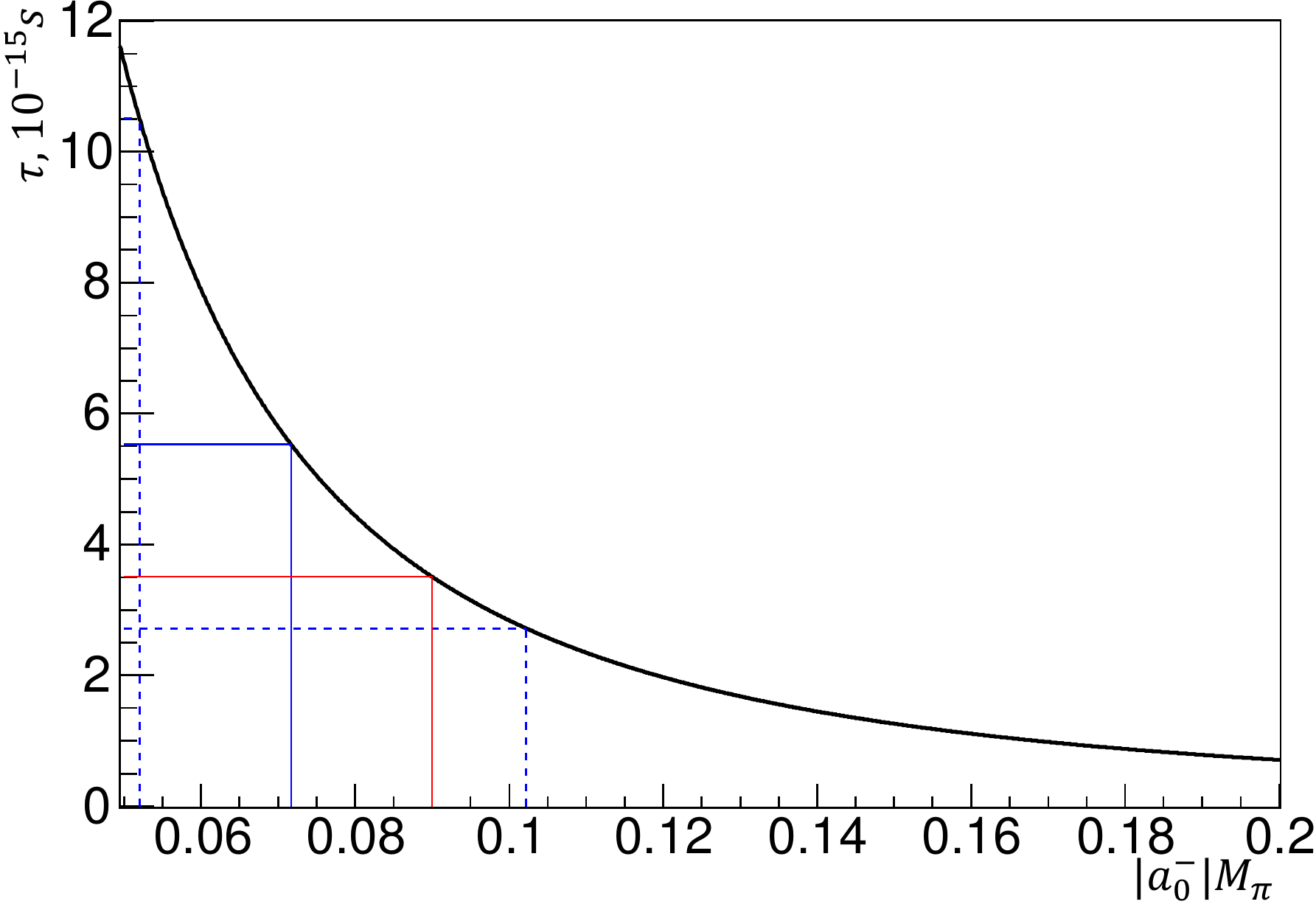}
\caption{Ground state $A_{\pi K}$ lifetime~$\tau_{1S}$ versus $a_0^-$ form
$Q$ analysis. Experimental results (blue lines) are compared to the theoretical 
prediction (red lines).}
\label{fig:tau_a0}
\end{figure}

The likelihood functions of the $(|Q_L|, Q_T)$ and $Q$ analyzes are shown 
in Fig.~\ref{fig:likelihoods}, and Table~\ref{tab:piKlifetime} summarizes 
the results of both analysis types and for different cuts in the $Q$ space. 
One realizes that the usage of the Pt data in the analysis does not 
significantly modify the final result. As the magnitude of the systematic error 
for Pt is only about 2 times smaller than the statistical uncertainty, 
the inclusion of systematic errors changes the relative weights 
of the Pt and Ni data samples, thus shifting the best estimate 
for $\tau_{\text{tot}}$ with respect to $\tau_{\text{stat}}$. 
The introduction of the criteria  $|Q_x|,|Q_y|<4\:\text{MeV}/c$ increases 
the background level by 22\%, relative to the criterion $Q_T<4\:\text{MeV}/c$. 
The results in Table~\ref{tab:piKlifetime} show that the lifetime values 
obtained with the $Q$~analysis are practically equal for both criteria. 
Therefore, the final result is presented for the $Q$ analysis evaluated with 
the criterion $Q_T<4\:\text{MeV}/c$, using the statistics of the Ni and 
Pt data samples:
\begin{equation}
\tau_{\text{tot}} = \left.
(5.5^{+5.0}_{-2.8}\right|_{\text{tot}})\cdot10^{-15}~\text{s}.
 \label{eq:piKlifetime_Q}
\end{equation}
The measured $\pi K$ atom lifetime corresponds, according to 
the relation (\ref{eq:gamma}) (Fig.~\ref{fig:tau_a0}), to 
the following value of the $\pi K$  scattering length $a_0^-$: 
\begin{equation}
\left|a_0^-\right| M_{\pi} = \left. 0.072^{+0.031}_{-0.020}\right|_{\text{tot}}.
\end{equation}

\begin{table}
\caption{$\pi K$~atom lifetime measurements:
$\tau_{\text{stat}}$ (only statistical error) and 
$\tau_{\text{tot}}$ (total error) in $10^{-15}$~s.}
\label{tab:piKlifetime}
\begin{ruledtabular}
\begin{tabular}{ccccc}
Analysis & Cuts & Target & $\tau_{\text{stat}}$ & $\tau_{\text{tot}}$ \\ \hline
$(|Q_L|, Q_T)$ & $Q_T<4\:\text{MeV}/c$ & Pt\&Ni  &
$3.96^{+3.49}_{-2.12}$ & $3.79^{+3.48}_{-2.12}$ \\
$(|Q_L|, Q_T)$ & $Q_T<4\:\text{MeV}/c$ & Ni  &
$3.52^{+3.40}_{-2.10}$ & $3.52^{+3.42}_{-2.11}$ \\
$(|Q_L|, Q_T)$ & $|Q_x|,|Q_y|<4\text{MeV}/c$ & Pt\&Ni  &
$3.16^{+2.67}_{-1.73}$ & $2.89^{+2.63}_{-1.70}$ \\
$(|Q_L|, Q_T)$ & $|Q_x|,|Q_y|<4\text{MeV}/c$ & Ni  &
$2.66^{+2.56}_{-1.66}$ & $2.66^{+2.58}_{-1.66}$ \\ \hline
$Q$ & $Q_T<4\:\text{MeV}/c$ & Pt\&Ni &
$5.64^{+4.99}_{-2.82}$ & $5.53^{+4.98}_{-2.81}$ \\
$Q$ & $Q_T<4\:\text{MeV}/c$ & Ni &
$5.07^{+4.73}_{-2.74}$ & $5.07^{+4.77}_{-2.75}$ \\    
$Q$ & $|Q_x|,|Q_y|<4\text{MeV}/c$ & Pt\&Ni &
$5.62^{+4.65}_{-2.71}$ & $5.60^{+4.68}_{-2.72}$ \\
$Q$ & $|Q_x|,|Q_y|<4\text{MeV}/c$ & Ni &
$4.98^{+4.37}_{-2.60}$ & $4.98^{+4.41}_{-2.62}$ \\
\end{tabular}
\end{ruledtabular}
\end{table}

All theoretical predictions are compatible with the measured value taking into
account the experimental precision. The main contribution to the experimental
uncertainty comes from statistics. As shown in \cite{GORC16}, the number of 
$\pi K$ atoms detected per time unit would be increased by a factor of 30 to 
40, if the DIRAC experiment could exploit the CERN SPS 450~$\text{GeV}/c$ 
proton beam.Under these conditions, the statistical precision of $a_0^-$ will 
be around 5\% for a single run period.

\section{Conclusion} 
\label{sec:concl}

The DIRAC Collaboration published the observation of 
$\pi^- K^+$ and $\pi^+ K^-$ atoms \cite{ADEV16}. These atoms were generated 
by the 24~$\text{GeV}/c$ protons of the CERN PS in Ni and Pt targets, 
where a part of them broke up, yielding $\pi^- K^+$ and 
$\pi^+ K^-$ atomic pairs. In the present article, 
the breakup probabilities for each atom type and each target 
are determined by analyzing atomic and free $\pi K$ pairs. 
By means of these probabilities, the lifetime of the $\pi K$ atom in
the ground state is evaluated, 
$\tau_{\text{tot}} = 
\left.(5.5^{+5.0}_{-2.8}\right|_{\text{tot}})\cdot10^{-15}$~s, 
and the S-wave isospin-odd $\pi K$ scattering length deduced, 
$\left|a_0^-\right| = 
\frac{1}{3}\left|a_{1/2}-a_{3/2}\right| = \left.
(0.072^{+0.031}_{-0.020}\right|_{\text{tot}}) M_{\pi}^{-1}$. 
The measured $a_0^-$ value is compatible with 
our previous less precise result \cite{ADEV14} 
and with theoretical results calculated in ChPT, LQCD 
and in a dispersive framework using Roy-Steiner equations 
\cite{SCHW04,BUET04,WEIN66,Gasser85,BERN91,KUBI02,BIJN04,JANO14,
BEAN06,Fu12,Sasaki14}.

On the basis of the statistically significant observation of 
$\pi K$ atoms \cite{ADEV16}, DIRAC presents 
a measurement of the $\pi K$ atom lifetime and the corresponding 
fundamental $\pi K$ scattering length.

\section*{Acknowledgements}

We are grateful to R.~Steerenberg and the CERN PS crew for the
delivery of a high quality proton beam and the permanent effort to
improve the beam characteristics. We thank G.~Colangelo,
J.~Gasser, H.~Leutwyler, U.G. Meissner, B.~Kubis, A.~Rusetsky,
M.~Ivanov and O.~Teryaev for their interest to our work and helpful
discussions. The project DIRAC has been supported by CERN and JINR
administrations, Ministry of Education and Youth of the Czech Republic
by project LG130131, the Istituto Nazionale di Fisica Nucleare and the
University of Messina (Italy), the Grant-in-Aid for Scientific
Research from the Japan Society for the Promotion of Science, the
Ministry of Education and Research (Romania), the Ministry of
Education and Science of the Russian Federation and the Russian Foundation
for Basic Research, the Direcci\'{o}n Xeral de Investigaci\'{o}n,
Desenvolvemento e Innovaci\'{o}n, Xunta de Galicia (Spain) and the
Swiss National Science Foundation.

\end{document}